\documentclass[a4paper,fleqn,usenatbib]{mnras}

\usepackage{mathptmx}

\usepackage[T1]{fontenc}
\usepackage{ae,aecompl}

\usepackage{graphicx}	
\usepackage{amsmath}	
\usepackage{amssymb}	

\newcommand{\Chandra}{\textit{Chandra}}

\newcommand{\NH}{\ensuremath{N_{\rm H}}}
\newcommand{\NHtwo}{\ensuremath{N_{\rm H,22}}}
\newcommand{\Ts}{\ensuremath{T_{\rm s}}}
\newcommand{\Tb}{\ensuremath{T_{\rm b}}}
\newcommand{\Msun}{\ensuremath{M_{\rm Sun}}}
\newcommand{\rhob}{\ensuremath{\rho_{\rm b}}}
\newcommand{\Tssix}{\ensuremath{T_{\rm s,6}}}
\newcommand{\Tc}{\ensuremath{T_{\rm c}}}

\title[DNB in NS cooling and Cas A]{Diffusive nuclear burning in cooling simulations and application to new temperature data of the Cassiopeia A neutron star}
\author[M.J.P. Wijngaarden et al.]{
M. J. P. Wijngaarden,$^{1}$\thanks{E-mail: M.J.P.Wijngaarden@soton.ac.uk}
Wynn C. G. Ho$^{1,2,3}$,
Philip Chang$^{4}$,
Craig O. Heinke$^{5}$,
\newauthor
Dany Page$^{6}$,
Mikhail Beznogov$^{6}$
and Daniel J. Patnaude$^{7}$
\\
$^{1}$Mathematical Sciences and STAG Research Centre, University of Southampton, SO17 1BJ, Southampton, UK\\
$^{2}$Physics and Astronomy, University of Southampton, SO17 1BJ, Southampton, UK\\
$^{3}$Department of Physics and Astronomy, Haverford College, 370 Lancaster Avenue, Haverford, PA, 19041, USA\\
$^{4}$Department of Physics, University of Wisconsin-Milwaukee, 1900 E. Kenwood Blvd., Milwaukee, Wisconsin 53211, USA\\
$^{5}$Department of Physics, University of Alberta, CCIS 4-181, T6G 2E1, Edmonton, Alberta, Canada\\
$^{6}$Instituto de Astronom\'ia, Universidad Nacional Aut\'onoma de M\'exico, Mexico City, CDMX 04510, Mexico\\
$^{7}$Smithsonian Astrophysical Observatory, Cambridge, MA 02138, USA
}

\date{Accepted 2018 December 29. Received 2018 December 11; in original form 2018 October 25}

\pubyear{2018}

\begin{document}
\label{firstpage}
\pagerange{\pageref{firstpage}--\pageref{lastpage}}
\maketitle

\begin{abstract}
	
A critical relation in the study of neutron star cooling is the one between surface
temperature and interior temperature. This relation is determined by the composition
of the neutron star envelope and can be affected by the process of diffusive nuclear
burning (DNB), which occurs when elements diffuse to depths where the density and
temperature are sufficiently high to ignite nuclear burning. We calculate models of
H-He and He-C envelopes that include DNB and obtain analytic temperature relations
that can be used in neutron star cooling simulations. We find that DNB can lead to a
rapidly changing envelope composition and prevents the build-up of thermally stable
hydrogen columns $y_H \gtrsim 10^{7}$ g cm$^{-2}$, while DNB can make helium envelopes more
transparent to heat flux for surface temperatures $\Ts \gtrsim 2\times$ 10$^6$ K.  We perform
neutron star cooling simulations in which we evolve temperature and envelope
composition, with the latter due to DNB and accretion from the interstellar medium.
We find that a time-dependent envelope composition can be relevant for understanding
the long-term cooling behaviour of isolated neutron stars. We also report on the
latest \textit{Chandra} observations of the young neutron star in the Cassiopeia A supernova
remnant; the resulting 13 temperature measurements over more than 18 years yield
a ten-year cooling rate of $\approx$ 2\%. Finally, we fit the observed cooling trend of
the Cassiopeia A neutron star with a model that includes DNB in the envelope.

\end{abstract}

\begin{keywords}
dense matter -- diffusion -- stars: evolution -- stars: neutron -- supernovae: individual: Cassiopeia A -- X-rays: stars.
\end{keywords}


\section{Introduction}
\label{sec:intro}

The physical properties, such as composition and structure, of neutron stars (NSs) are still uncertain. Valuable information about the neutron star interior can be obtained by comparing observations of thermal radiation of cooling NSs with theoretical models of how a NS cools over time. Isolated NSs cool via neutrino and photon emission after they are formed hot in a supernova explosion (see review by \citealt{2015SSRv..191..239P}). Additionally, cooling models can be used to study the cooling of accretion-heated NSs in binary systems (for a recent review see \citealt{2017JApA...38...49W}). 

Observations of several isolated neutron stars suggest that the envelopes of younger pulsars ($<$10$^{4}-$10$^{5}$ yr) may consist of light elements such as hydrogen or helium, whereas the envelopes of older pulsars ($>$10$^5$ yr) may be composed of heavier elements (see, e.g., \citealt{2001A&A...379L...5Y,2014PhyU...57..735P}). Evidence for the presence of carbon in a NS envelope was found for the first time using spectral observations of the young NS in Cassiopeia A \citep{2009Natur.462...71H}. Since then, several other NS spectra are found to be fit by carbon atmosphere spectra (see \citealt{2017JPhCS.932a2006D}, and references therein). These observations indicate a possible evolution of the composition of the envelope through nuclear burning, after the initial composition is set by fallback material onto the NS after a supernova explosion. The NS in Cassiopeia A (we hereafter refer to the NS as Cas A) is an intriguing object not only because of the chemical composition of its envelope, but also because surface temperatures derived from X-ray observations suggest that the source might be cooling rapidly (\citealt{2010ApJ...719L.167H}; c.f. \citealt{2013ApJ...779..186P}). Constraining cooling models with multiple temperature observations can provide fundamental knowledge on the properties of the NS interior, such as critical temperatures for superconductivity and superfluidity (e.g., \citealt{2011MNRAS.412L.108S,2011PhRvL.106h1101P,2015PhRvC..91a5806H}). Other possibilities include fast or rotationally induced neutrino cooling \citep{2013PhLB..718.1176N,2016MNRAS.456.1451T}, magnetic field decay \citep{2014A&A...561L...5B}, slow thermal relaxation \citep{2012PhRvC..85b2802B,2013PhRvC..88f5805B}, stellar fluid oscillations \citep{2011ApJ...735L..29Y}, and transition to axions or quark matter \citep{2013ApJ...765....1N,2013A&A...555L..10S,2014JCAP...08..031L,2018PhRvD..98j3015H}.

A major source of uncertainty in cooling models is the chemical composition of the thin outer envelope which extends from just below the surface down to a boundary mass density of $\rho_b \sim 10^{8} - 10^{10}$~g~cm$^{-3}$. The envelope acts as a thermal insulator due to its relatively poor thermal conductivity and thus sets a strong temperature gradient between the surface temperature T$_s$ and the interior temperature at the bottom of the envelope T$_b$ [$\equiv T(\rho_b)$]. The composition affects the thermal conductivity and thus how transparent the envelope is to the heat flux. This means that a measured surface temperature can correspond to different interior temperatures depending on the composition of the envelope. 

Several models of the heat blanketing envelope have been developed after the work by \cite{1983ApJ...272..286G}, who calculated the thermal structure and conductivity of an iron envelope. Later models went beyond the single element envelope and considered the envelope as multiple shells of light elements (H, He, C) separated by an abrupt boundary \citep{1997A&A...323..415P,2003ApJ...594..404P}. These models have subsequently been used in many interior studies. Most recently, \cite{2016MNRAS.459.1569B} developed new models for binary ion mixtures (H-He, He-C and C-Fe) where ion diffusion is taken into account but the possible effect of nuclear burning due to diffusion is not included.

Numerical calculations of the neutron star envelope including the effects of diffusive nuclear burning (DNB) have been performed by \cite{2003ApJ...585..464C,2004ApJ...605..830C,2010ApJ...723..719C}. The key idea of DNB is that the tail of the light element layer may diffuse down to depths where the density and temperature is large enough to ignite burning. The depletion of the light elements at this depth can subsequently drive a light element current from the surface to the burning layer. In \cite{2003ApJ...585..464C} and \cite{2004ApJ...605..830C}, a H-C envelope was modelled, and it was found that diffusive nuclear burning can readily consume all of the initial hydrogen in the NS envelope within $\sim$10$^5$ yr. Motivated by the detection of carbon on the surface of Cas A, \cite{2010ApJ...723..719C} calculated models for DNB in He-C envelopes and found that any primordial helium is likely depleted during the hot ($T_b$> 10$^8$ K), early evolution of the NS.

Typically, NS cooling models calculate the NS structure and thermal evolution from the centre out to the bottom of the envelope, where they are coupled to theoretical envelope temperature relations (between T$_b$ and T$_s$) based on the envelope models above. Diffusive nuclear burning can change the composition of the envelope and hence its thermal conductivity. Therefore, it can affect the temperature relations between T$_s$ and T$_b$. We use the work of \cite{2003ApJ...585..464C,2004ApJ...605..830C,2010ApJ...723..719C} and calculate models including diffusive nuclear burning for H-He and He-C envelopes and provide analytic envelope relations (based on those presented in \citealt{2016MNRAS.459.1569B}) that can be used in cooling models. We use these temperature relations to investigate the effect of diffusive nuclear burning and time-variable envelopes on neutron star cooling. We also report on four new \textit{Chandra} observations of the NS in Cassiopeia A resulting in the longest spanning ( $>$18 years) dataset currently available. We then use the new temperature relations including DNB to model the observed cooling of Cas~A.

\section{Theory of diffusive nuclear burning}
\label{sec:theory}

\subsection{The physics of DNB}
\label{sec:dnbphysics}

We summarize the key physics input for diffusive nuclear burning as calculated in \cite{2010ApJ...723..719C}. For a full description of the physics we refer the reader to this work. The neutron star envelope, at $\rho$ < 10$^{8}- $10$^{10}$ g cm$^{-3}$, has a thickness ($\sim$ a few hundred metres) that is much smaller than the stellar radius and can be well approximated as a plane-parallel layer with a constant downward gravitational acceleration $g$. The thermal profile is obtained by solving the heat diffusion equation for a constant flux in radiative equilibrium

\begin{equation}
\frac{\partial T}{\partial z} = - \frac{3 \kappa \rho}{16T^3} T_{s}^4,
\end{equation}

\noindent
where $z$ is the depth, $T_s$ is the effective (non-redshifted) surface temperature and $\kappa$ is the opacity. In the outermost layer of the envelope the radiative opacity is dominated by free-free absorption and Thomson scattering. We use the radiative opacities of \cite{2001A&A...374..213P} and the analytic formulae for the conductive opacity given by \cite{1999A&A...346..345P}.

We consider the case where a trace ion species, hereafter referred to as `light elements',  is immersed in a fixed background `heavy' ion species. The equation for hydrostatic equilibrium, which is dominated by electric and gravitational forces (see \citealt{2010ApJ...723..719C}) is used to obtain the number density profiles of ion species in the absence of nuclear burning. We require overall charge neutrality such that, $n_{e} ~=~ Z_{\text{light}} n_{\text{light}} + Z_{\text{heavy}} n_{\text{heavy}}$, where $n_e$ is the electron number density, and $Z_{\text{light}}$,  $n_{\text{light}}$, $Z_{\text{heavy}}$ and $n_{\text{heavy}}$ are the charge number and number density of the light (trace) and heavy (background) element species, respectively. The diffusive tail of the light element layer may extend deep down into the underlying heavier element layer. When lighter elements diffuse to depths where they are consumed by nuclear burning, the depletion of light elements at this depth induces a diffusive light element current from the surface to the burning layer. Because the timescale at which the local column density changes is much longer than the local nuclear timescale, a steady state approximation is used for the light element current induced by DNB ($\partial n_{\text{light}}/ \partial t = 0$). The light element current is defined as

\begin{equation}
J_{\text{light}} = - \mathcal{D} \frac{dn_{\text{light}}}{dz} + n_{\text{light}} w_{\text{light}} = n_{\text{light}} v_{\text{light}},
\end{equation}

\noindent
where $v_{\text{light}}$ and $w_{\text{light}}$ are the light element relative velocity and drift velocity, respectively. The diffusion coefficient $\mathcal{D}$ is calculated as in \cite{2004ApJ...605..830C} (similar to \citealt{2002ApJ...574..920B})

\begin{equation}
\mathcal{D} \approx 10^{-3} \frac{A_{\text{light}}^{0.1} T_6^{1.3}}{Z_{\text{light}}^{1.3} Z_{\text{heavy}}^{0.3} \rho_5^{0.6}} \mathrm{cm}^2 \mathrm{s}^{-1},
\end{equation}

\noindent
where $T_6$ = $T/$10$^6$ K and $\rho_5$ = $\rho$/10$^5$ g cm$^{-3}$. The atomic mass and charge number are denoted by $A$ and $Z$, respectively. From the light element continuity equation and the definition of the current, one can derive the general case of the steady-state diffusion equation (which is derived in detail in \citealt{2004ApJ...605..830C})

\begin{equation}
\label{eq:diffion}
\frac{\partial^2 f_{\text{light}}}{\partial z^2} + \frac{\partial f_{\text{light}}}{\partial z} \frac{\partial \text{ln}~n_{\text{light},0}}{\partial z} = \frac{f_{\text{light}}}{\mathcal{D} \tau_{\text{light}}},
\end{equation}

\noindent
where $n_{\text{light},0}$ is the number density of the trace ion in the absence of nuclear burning obtained by solving the hydrostatic balance equations numerically, $f_{\text{light}}$ is the correction factor to the number density due to nuclear burning (i.e., $f_{\text{light}} \equiv n_{\text{light}}/n_{\text{light},0}$) and $\tau_{\text{light}}$ is the local lifetime against nuclear capture set by the local nuclear burning rate. With increasing depth and temperature, the ion scale height (=$[d$ln~$n_{\text{light,0}}/dz$$]^{-1}$) rapidly becomes larger than the nuclear scale height (=$\sqrt{\mathcal{D}\tau_{\text{light}}})$. Therefore, at the burning layer where the second term can effectively be dropped, Equation \ref{eq:diffion} represents a diffusion equation for $f_{\text{light}}$ with a nuclear-driven source.

In our models, hydrogen can be consumed steadily by the proton capture reactions of the CNO-cycle and pp chain reactions. Helium is consumed by proton- and alpha capture reactions and the triple alpha process, although the reaction rates of the triple alpha process were found to be negligible in the parameter space of interest (see \citealt{2010ApJ...723..719C}) and thus they are ignored. Nuclear reaction rates and experimental reaction cross section values are obtained from the NACRE\footnote{\url{http://pntpm.ulb.ac.be/Nacre/nacre.htm}} \citep{1999NuPhA.656....3A} and REACLIB\footnote{\url{http://nucastro.org/reaclib.html}} database compilations. 

\subsection{Model description}
\label{sec:model}

We model the neutron star envelope for $^1$H-$^4$He and $^4$He-$^{12}$C mixtures following the work of \cite{2016MNRAS.459.1569B}, but including the physics of DNB. These envelope compositions were chosen to be able to compare our solutions with those found in \cite{2016MNRAS.459.1569B}, but in reality other ions may be present in neutron star envelopes. In particular, one might consider a H-He-C envelope, with an intervening helium layer between the hydrogen and carbon components. The effects of such an envelope are strongly dependent on the assumed size of each layer.  For simplicity in the present work, we limit our study of DNB to two component envelopes. We use a surface gravity of $g_{s}$ =  2.43 $\times$ 10$^{14}$ cm s$^{-2}$ in our calculations, but the resulting temperature relations at other surface gravities can be approximated using the temperature scaling relation by \cite{1983ApJ...272..286G}.

We start our numerical calculations with a given surface temperature $T_s$ and fraction $x_{\text{light}}$ of the light element species present at the photosphere. The partial pressure $p_{\text{light}}$ and column density $y_{\text{light}}$ of the ion components at the photosphere is set by $x_{\text{light}}$ as y$_{col}$ = $y_{\text{light}} + y_{\text{bkg}}$ = $x_{\text{light}} \times y_{col} + x_{\text{bkg}} \times y_{col}$. We then integrate the system of equations in Section \ref{sec:dnbphysics} to obtain profiles for the internal temperature, ion densities and nuclear burning rates. A given photospheric light element fraction can correspond to different values of the total light element column density depending on the surface temperature. Therefore, we perform a large number of numerical calculations for varying $T_s$ and $x_{\text{light}}$ to obtain sufficient coverage in the resulting $T_s$-$T_b$-$y_{\text{light}}$ parameter space that is used for fitting analytic functions. 

We note that in the analytic temperature relations calculated by \cite{2016MNRAS.459.1569B}, the amount of light elements is characterized by their effective transition density $\rho*$. This is an artificially chosen density at which a total mass of light elements is contained in the outer shell $\rho < \rho*$. Note that the light element column density $y_\text{light}$ can be converted to $\rho*$ making use of the relevant equation of state. 

\subsection{Parameter ranges}

We fit the numerical results of the models described above to obtain analytic $T_s-T_b$ relations that can be coupled to thermal evolution codes for the neutron star interior. Note that the $T_s-T_b$ relations are actually relations for $T_s$, $T_b$ and $y_{\text{light}}$ (or equivalently, $\rho*$) for a given surface gravity, i.e. T$_b$($T_s$, $y_{\text{light}}$). We are limited in our parameter ranges by both physical and numerical constraints. Physical constraints arise for example from the fact that the light element column density cannot be larger than the envelope-crust boundary column density $y_b$ or smaller than the photospheric density. For the H-He envelopes, we let $\rho_b$ = 10$^8$ g cm$^{-3}$ and increase $\rho^*$ up to $\sim$10$^7$ g cm$^{-3}$. For He-C envelopes we use $\rho_b$ = 10$^{10}$ g cm$^{-3}$ and let $\rho^*$ vary up to $\sim$10$^{8}$ g cm$^{-3}$. The surface temperature range for all compositions is $0.3 \times 10^{6}$ K$< T_s < 3 \times 10^{6}$ K, such that it encompasses the observed temperatures for both isolated neutron stars and transient neutron stars in quiescence.

\section{Burning rates}
\label{sec:burning}

The reaction rates due to DNB depend on the composition of the envelope, the amount of burning material available and the temperature in the burning region. Using the nuclear reactions that consume hydrogen and helium discussed in Section \ref{sec:theory}, the total burning rate is

\begin{equation}
\label{eq:zeta}
\zeta_{{\rm DNB}} = \frac{y_{\text{light}}}{\tau_{\text{light,col}}} = \int dz \frac{A_{\text{light}} n_{\text{light}} m_p}{\tau_{\text{light}}(n_{\text{light}},n_{\text{sub}},T)},
\end{equation}

\noindent
where $A_{\text{light}}$ is the atomic number of the light element species, $\tau_{\text{light,col}}$ is the characteristic time for the total integrated light element column to be consumed and $\tau_{\text{light}}$ is the local lifetime of the light element species due to DNB. The local number densities of the $\alpha$- or $p$-capturing substrate are denoted by $n_{\text{sub}}$.

\begin{figure}
	\includegraphics[width=\columnwidth]{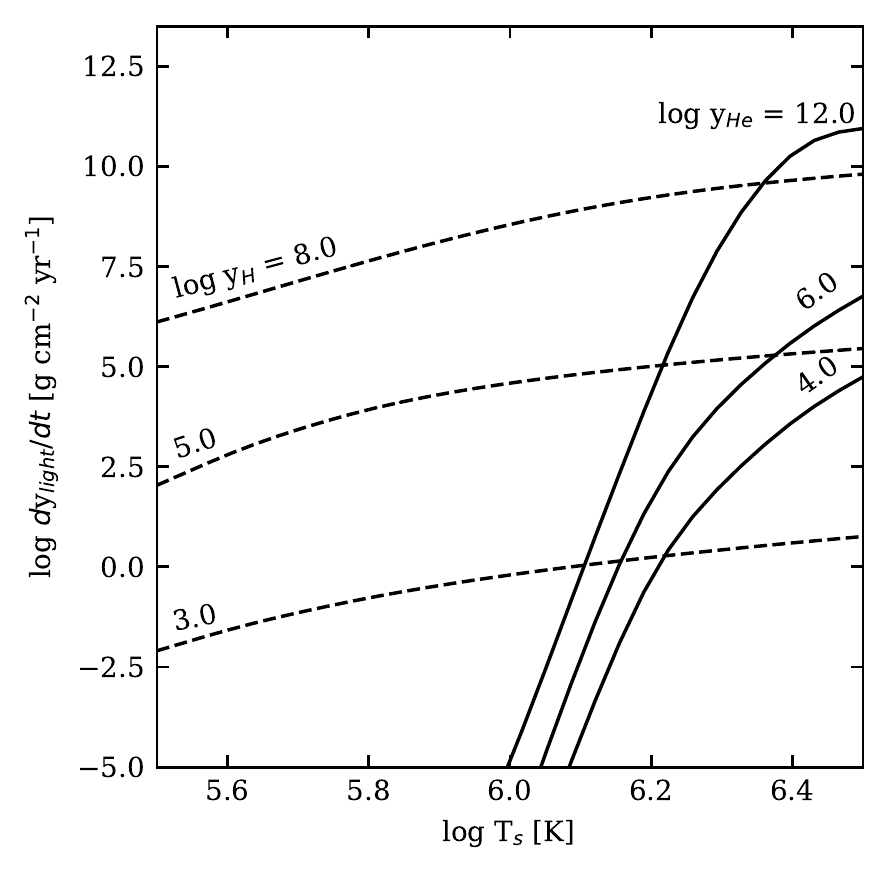}
	\vspace{-20pt}
	\caption{Light element column density burning rates for hydrogen burning (dashed) and helium burning (solid). For each line the value of log y$_{\rm light}$ is indicated. Note that $d$log $y_{\rm light}/dt$ is the same as log $\zeta_{\rm DNB}$; see Equation \ref{eq:zeta}.}
	\label{fig:burningrates}
\end{figure}

In Figure \ref{fig:burningrates} we plot the resulting burning rates for several values of y$_{\rm light}$ for both the H-He and He-C envelope models. For hydrogen burning (dashed lines in Figure \ref{fig:burningrates}), we see that the burning rates are more sensitive to the amount of material available for burning (as characterised by the hydrogen column density) than to the temperature (characterised by the surface temperature). The strong dependence on column density is not surprising, as the burning region is located in the exponentially suppressed diffusive hydrogen tail. Interestingly, Figure \ref{fig:burningrates} shows that DNB can alter the hydrogen column density significantly for the entire range of plotted T$_s$. Especially for intermediate and large hydrogen column densities (y$_{\rm H} \gtrsim$ 10$^5$ g cm$^{-2}$) the hydrogen column density lifetime is $<<$ 10$^4$ yr. It is clear that the helium burning rates drop off rapidly with decreasing temperatures, which is expected as the dominant burning processes are the highly temperature sensitive $\alpha$-capturing reactions.  We find that the helium burning rate effectively switches off for T$_s~\lesssim$ 10$^6$ K. At temperatures T$_s~\gtrsim$ 10$^6$ K the burning rate is sufficiently large to change the helium column density within years.

\begin{figure}
	\includegraphics[width=\columnwidth]{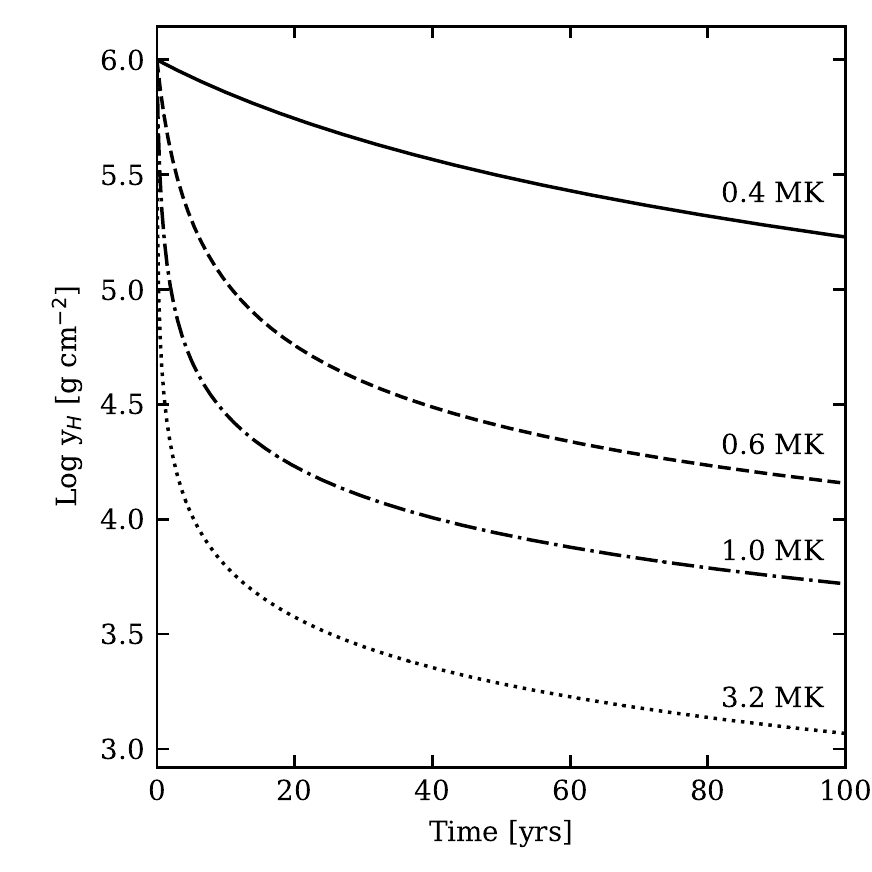}
	\vspace{-20pt}
	\caption{Evolution of the hydrogen column density for a constant surface temperature [(0.4, 0.6, 1.0, 3.2)$\times$ 10$^6$  K] and initial y$_{\rm H}$ = 10$^6$ g cm$^{-2}$. }
	\label{fig:Hevolution}
\end{figure}

\begin{figure}
	\includegraphics[width=\columnwidth]{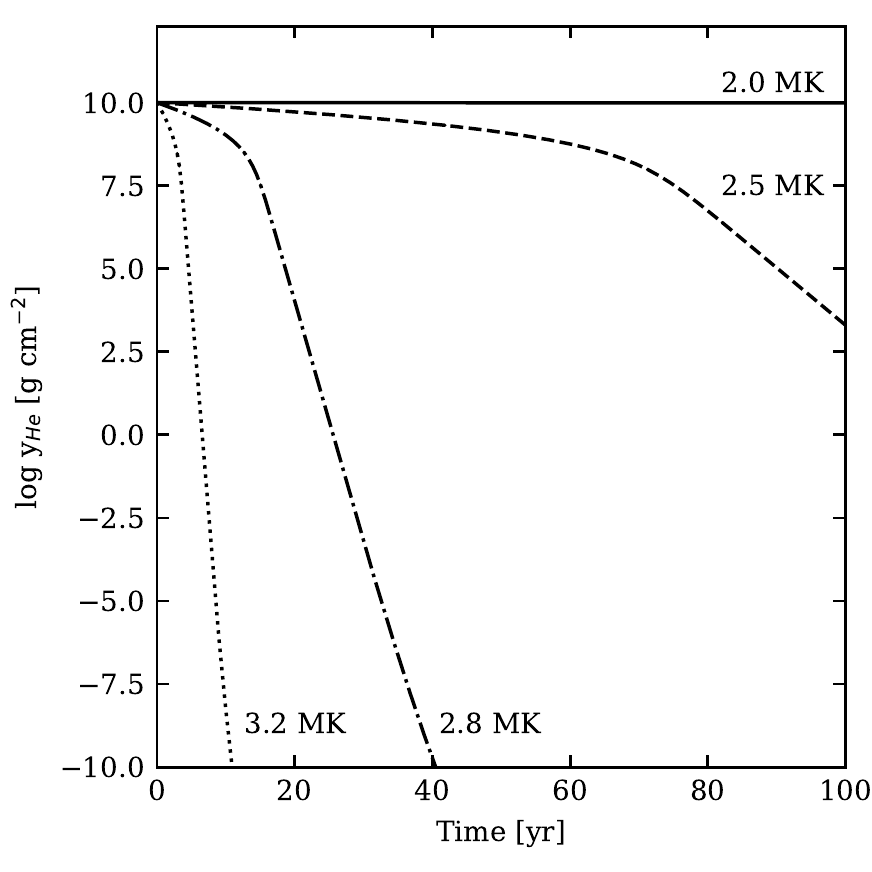}
	\vspace{-20pt}
	\caption{Evolution of the helium column density for a constant surface temperature [(2, 2.5, 2.8, 3.2)$\times$ 10$^6$ K] and initial y$_{\rm He}$ = 10$^{10}$ g cm$^{-2}$. }
	\label{fig:HeCevolution}
\end{figure}

Using the burning rates calculated above, we illustrate the evolution of the light element column over time for a fixed surface temperature in Figure \ref{fig:Hevolution} and \ref{fig:HeCevolution} for hydrogen and helium, respectively. We use an initial hydrogen column density of y$_{\rm light}$ = 10$^6$ g cm$^{-2}$ and vary the surface temperature between $ 0.4 \times 10^{6}<$ $\Ts < 3.2 \times 10^{6}$ K. For all temperatures, the sharpest drop in hydrogen column density occurs within the first $\sim$ 10 years due to the strong density dependence of the burning rate. After that time, the column decreases more gradually. For helium burning we use an initial helium column density of y$_{\rm light}$ = 10$^{10}$ g cm$^{-2}$ and vary the surface temperature between~$2 \times 10^{6}<$ $\Ts < 3.2 \times 10^{6}$ K. It is clear that the helium column change is more strongly coupled to the temperature and that for $\Ts \lesssim$  10$^{6}$ K the column does not change significantly within 100 years. At higher temperatures, Figure \ref{fig:HeCevolution} illustrates that DNB can even substantially alter the helium column on relatively short timescales of $< 10-100$ years.

\section{Static envelope temperature relations}
\label{sec:tbts}

We present analytic temperature relations that include diffusive nuclear burning in the envelope. The fit functions and parameters for each of the mixtures are given in Appendix \ref{sec:appendix_tbts}. In this section, we review their main characteristics and implications.

The temperature relations are most sensitive to the composition in a small region in the envelope which is commonly referred to as the sensitivity strip. The location of the sensitivity strip in the envelope is determined by where the radiative $\kappa_{r}$ and conductive $\kappa_{c}$ opacity are comparable, i.e., $\kappa_{r} \approx \kappa_c$. Even a small amount of light elements in this region can have a large effect on the resulting temperature profile in the envelope. The composition outside of the sensitivity strip barely affects the temperature relation. Therefore, in order to understand the effect that DNB has on the temperature relations, $\Tb(\Ts, y_{\rm light})$, we will consider for which values of $y_{\rm light}$ the composition in the sensitivity strip is altered by DNB.
	
\begin{figure}
	\includegraphics[width=\columnwidth]{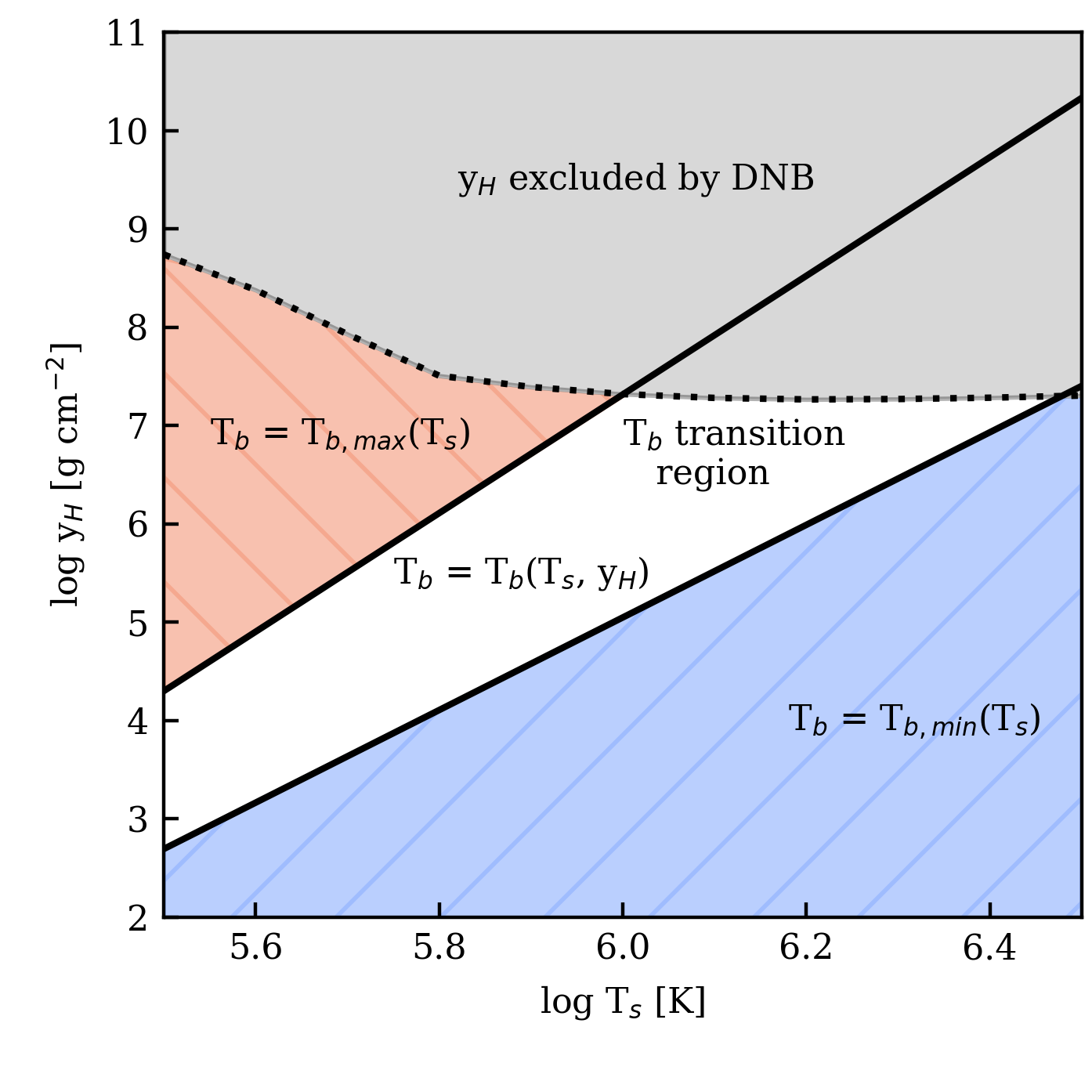} 
	\vspace{-20pt}
	\caption{$T_s$-$y_{\rm H}$ parameter space for the H-He envelope model with $\rho_b$ = 10$^8$ g cm$^{-3}$. The solid lines enclose the region where the boundary temperature is sensitive to the hydrogen column density, i.e. $\Tb$~=~$\Tb(T_s,y_{\rm H})$. In this region, the boundary temperature transitions from that corresponding to a pure hydrogen envelope (red), where $\Tb$~=~$T_{\text{b,max}}(\Ts)$, to that of a pure helium envelope (blue), where $\Tb$~=~$T_{\text{b,min}}(\Ts)$. Hydrogen column densities outside of the transition region have negligible effect on the $\Tb$-$\Ts$ relations. The grey shaded region shows the parameter space that is excluded due to DNB (see text). Note that at large $\Ts$, some excluded hydrogen columns overlap with the transition region. This means that the spread in $\Tb$ is smaller when DNB is taken into account.}
	\label{fig:sensitivity2}
\end{figure}

\begin{figure}
	\includegraphics[width=\columnwidth]{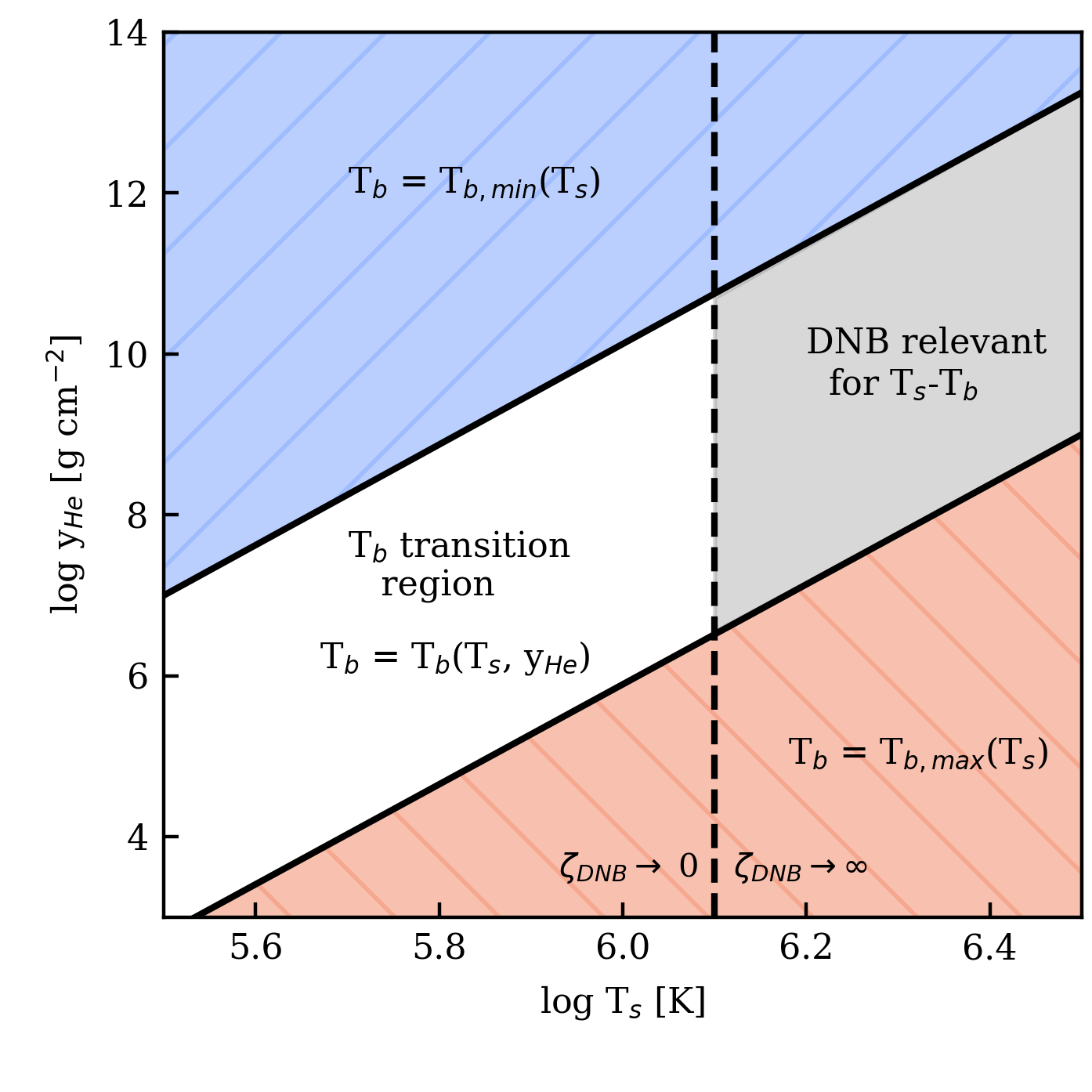} 
	\vspace{-20pt}
	\caption{$T_s$-$y_{\rm He}$ parameter space for a He-C envelope model with $\rho_b$ = 10$^{10}$ g cm$^{-3}$. The solid lines enclose the region where the boundary temperature is sensitive to the helium column density, i.e. $\Tb$~=~$\Tb(T_s,y_{\rm He})$. In this region, the boundary temperature transitions from that corresponding to a pure helium envelope (blue), where $\Tb$~=~$T_{\text{b,min}}(\Ts)$, to that of a pure carbon envelope (red), where $\Tb$~=~$T_{\text{b,max}}(\Ts)$. Helium column densities outside of the transition region have a negligible effect on the $\Tb$-$\Ts$ relations. The dashed line indicates the surface temperature above which the DNB rate $\zeta_{\rm DNB}$ is significant. For helium column densities in the transition region where $\zeta_{\rm DNB}$ is significant (highlighted in grey), DNB alters the $\Tb$-$\Ts$ relations considerably.  }
	\label{fig:sensitivity}
\end{figure}

In Figure \ref{fig:sensitivity2}, we show the $\Ts$-$y_{\rm H}$ parameter space for a H-He envelope; it is important to note that what is illustrated in Figures \ref{fig:sensitivity2} and \ref{fig:sensitivity} is light element column density, not total column density.  For small hydrogen columns and high surface temperatures (lower right blue region), the bulk of the hydrogen remains close to the surface and does not reach the sensitivity strip; thus the temperature relations resemble those of a pure helium envelope and are only a function of $\Ts$. The region enclosed by solid lines indicates temperatures and hydrogen columns which affect the composition of the sensitivity strip and thus the resulting $\Tb$. For large hydrogen columns (red region), hydrogen makes up most of the composition in the sensitivity strip, and increasing the column further does not change the composition inside the sensitivity strip. In this case, the temperature relations resemble a pure hydrogen envelope and are unaffected by the size of the hydrogen column. Hydrogen DNB occurs in the diffusive tail and the reaction rates are highly sensitive to the amount of material available at the burning depth (see Section \ref{sec:burning}). Therefore, DNB does not impact the composition distribution in the sensitivity strip for small hydrogen columns. For large H columns (y$_{\rm H} \gtrsim 10^{7}$ g cm$^{-2}$), the time for hydrogen to diffuse down to the burning depth is shorter than the rate at which it is consumed. In other words, the burning reactions are not limited by the diffusion time but by the nuclear reaction rate. Thus, hydrogen at the burning depth is consumed rapidly, which prevents the build up of large thermally stable hydrogen columns ($y_{\rm H}~\gtrsim$ 10$^{7}$~g~cm$^{-2}$). 

In Figure \ref{fig:sensitivity}, we show $\Ts$-$y_{\rm He}$ for the He-C envelope model. Diffusive helium burning takes place deep in the envelope (10$^{8}$~g~cm$^{-2}$~$\lesssim$~y~$\lesssim$~3~$\times$~10$^{12}$~g~cm$^{-2}$) and reaction rates are highly sensitive to the temperature (see Section \ref{sec:burning}). At $\Ts <$ 1.25 $\times$ 10$^{6}$ MK (denoted by the dashed line in Figure \ref{fig:sensitivity}), burning rates are too low to affect the envelope composition and temperature relations. Above this surface temperature, DNB prevents the helium column from extending deeper into the envelope beyond the burning depth. As a result, more helium resides at lower densities in the sensitivity strip compared to the case when DNB is not taken into account. Thus a He-C envelope is more  transparent to heat flux with DNB for sufficiently high temperatures.

Figures \ref{fig:sensitivity2} and \ref{fig:sensitivity} emphasise that the effect of hydrogen on the conductivity of the envelope is very different than that of helium, as first noted by \cite{2016MNRAS.459.1569B}. Typically, increasing the amount of light elements leads to a higher thermal conductivity of the envelope, such that for a given $\Ts$ the corresponding $\Tb$ is lower. An exception to this is the case of hydrogen, which has the opposite behaviour, as its larger radiative opacities results in a lower thermal conductivity. Therefore, increasing the amount of hydrogen, leads to a larger $\Tb$. 

\begin{figure}
	\includegraphics[width=\linewidth]{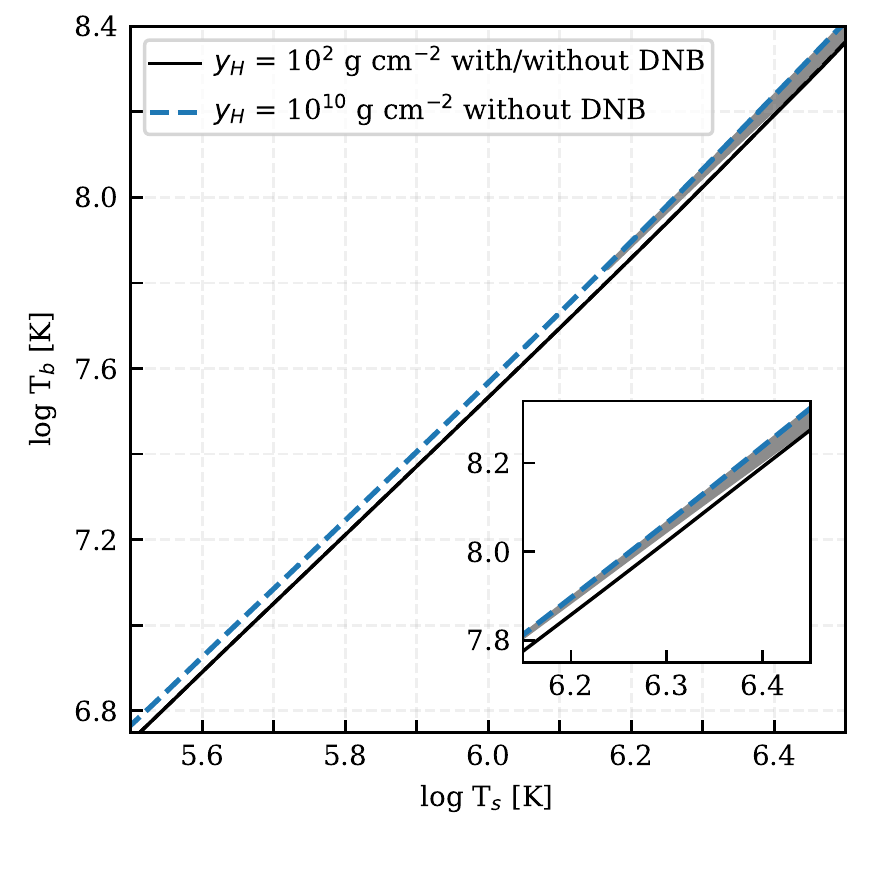}	
	\vspace{-20pt}
	\caption{The $T_s$-$T_b$ relation for the H-He envelope with $\rho_b$~=~10$^{8}$~g~cm$^{-3}$. The solid line corresponds to both the model with and without DNB. The blue dashed line corresponds to an hydrogen column of 10$^{10}$~g~cm$^{-2}$ when DNB is not included. The grey shaded region (see also inset) corresponds to the grey region in Figure \ref{fig:sensitivity2} and shows the region in the $\Ts-\Tb$ relations where DNB is reaction rate limited (see text). Note that the spread in $\Tb$ for a H-He envelope is small compared to other compositions and becomes even smaller at large temperatures due to DNB. }
	\label{fig:HHecomp}
\end{figure}

\begin{figure}
	\includegraphics[width=\linewidth]{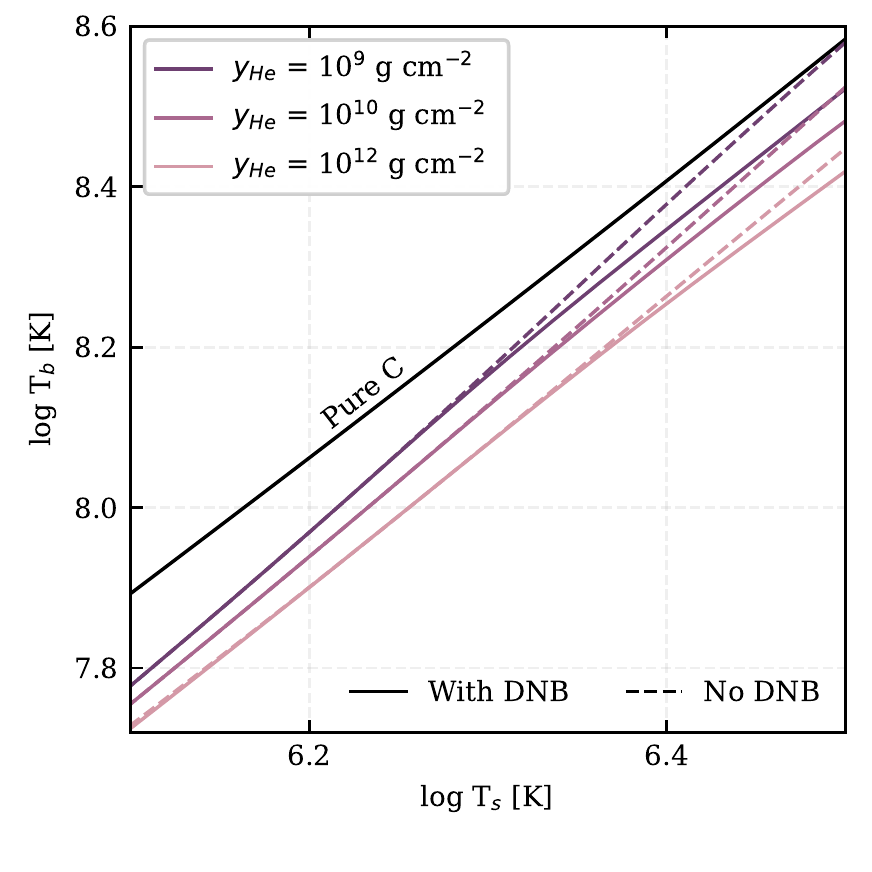}
	\vspace{-20pt}
	\caption{Comparison of the $T_s$-$T_b$ relation including DNB (solid) and without DNB (dashed) for the He-C envelope with $\rho_{\rm b}$~=~10$^{10}$~g~cm$^{-3}$. A small range in $\Ts$ is shown since DNB is negligible at lower $\Ts$ (see Figure \ref{fig:sensitivity}).}
	\label{fig:HeCcomp}
\end{figure}

The resulting $\Ts-\Tb$ relations for a H-He envelope including DNB are shown in Figure \ref{fig:HHecomp}. For comparison, the temperature relations where DNB is neglected are also plotted. It is clear that including DNB has a negligible effect on the thermal conductivity of the H-He envelope as the resulting temperature relations are indistinguishable. However, including DNB in the H-He envelope leads to an upper limit on the size of the hydrogen column, making the range in $\Tb$ for a given $\Ts$ smaller than when DNB is ignored (the difference is indicated by the shaded region in Figure \ref{fig:HHecomp}).

For a He-C envelope, the temperature relations are shown in Figure \ref{fig:HeCcomp}. As expected, these relations are unaffected by DNB at low temperatures.  At higher temperatures ($\Ts > 2 \times 10^{6}$ K; see Section \ref{sec:burning}) where helium burning is significant, DNB changes the temperature relations. In particular, for a given surface temperature, the boundary temperature is lower with DNB; equivalently, the same boundary temperature corresponds to a higher surface temperature. The He-C envelope is a better conductor and more transparent to the heat flux at large temperatures when DNB is taken into account since helium burning at large depths implies a greater helium abundance at shallower depths (in the sensitivity strip) for a given helium column (see above).

\section{Time variable envelopes}
\label{sec:timedependent}

Envelope temperature relations, such as the ones presented in Section \ref{sec:tbts}, are used to calculate the NS temperature evolution while assuming that the envelope composition does not change (see, e.g., \citealt{2015SSRv..191..239P} for a review of NS cooling).  The burning rates in Section \ref{sec:burning} indicate that modelling the envelope with a fixed composition over the course of the NS lifetime can be inaccurate.  In this section, we explore time-variable envelope compositions due to DNB and their effect on NS temperature evolution. Instead of calculating a cooling curve with a fixed envelope composition, we use a fixed initial composition and evolve both heat diffusion in the star and envelope composition.

\subsection{Cooling model}
\label{sec:coolmodel}

We model the thermal evolution of the NS using the general relativistic cooling code \textit{NSCool}\footnote{An older version of this code is available at \url{http://www.astroscu.unam.mx/neutrones/NSCool/}} \citep{2016ascl.soft09009P} for a NS with mass $M$~=~1.4~$\Msun$ and radius $R$~=~11.5 km, which is consistent with a surface gravity $g_{\mathrm s}=1.8\times 10^{14}\mbox{ cm s$^{-2}$}$. We use the nuclear A18+$\delta \nu$+UIX* equation of state \citep{1998PhRvC..58.1804A}. Neutron superfluidity is taken into account using the `SFB' $^1$S$_0$ gap model \cite{2003NuPhA.713..191S}, and proton superconductivity is modelled using the `CCDK' $^1S_0$ gap model \citep{1993NuPhA.555...59C}. Possible pairing of neutrons in the spin  triplet-state is not considered in this section but will be explored in the next one. While uncertainties in the superfluid properties of neutrons and superconducting properties of protons affect the cooling curves, we do not expect this to significantly change our results, as we use the same assumptions for all cooling curves and are only interested in the differences resulting from varying envelope compositions. The energy transport and conservation equations are solved up to the outer boundary defined as the bottom of the envelope: $\rhob=10^{8}\mbox{ g cm$^{-3}$}$ and $\rhob=10^{10}\mbox{ g cm$^{-3}$}$ for a H-He and He-C envelope, respectively. We use the $\Ts-\Tb$ models including DNB as described in Section \ref{sec:tbts} to relate the surface temperature and the boundary temperature.

\subsection{Variable envelope composition}
We use fits of the numerical burning rates to calculate the evolution of the light element column density $y_{\rm light}(t)$. In addition to solving equations describing the interior thermal profile, we evaluate the change in $y_{\rm light}$ at each time step. We model two competing processes that affect the evolution of column density. On the one hand, the amount of light elements decreases due to DNB when there is enough material available to burn and the temperature is sufficiently large. On the other hand, the amount of light elements increases due to accretion from the interstellar medium (ISM) or from a companion star.  A mass accretion rate $\dot{M}_{\text{acc}}$ is equivalent to an increasing light element column density $\dot{y}_{\text{acc}}$ via

\begin{equation}
\dot{y}_{\text{acc}} \approx \dot{M}_{\text{acc}} ~\frac{g_s}{4 \pi G M}.
\end{equation}

The total change in y$_{\text{light}}$ is obtained by combining the decreasing effect from DNB and the increasing effect from accretion (from the ISM). Thus the change in y$_{\text{light}}$ in a time interval $dt$ is

\begin{equation}
\Delta y_{\text{light}} = \frac{dy_{\text{light}}}{dt} dt = (~\dot{y}_{\text{acc}} - \zeta_{{\rm DNB}}~)~dt,
\end{equation}

\noindent
where $\zeta_{\rm DNB}$ is calculated as in Equation \ref{eq:zeta} for a grid of $\Ts-\Tb-y_{\rm light}$ and interpolated between gridpoints. Note that our steady state approximation is not strictly valid at early, hot stages immediately after NS birth, when the NS cools faster than diffusion through the envelope. A calculation of neutron star interior temperature evolution and diffusion in the
envelope is needed for a self-consistent solution, which is beyond the scope of this work.

\subsection{ISM accretion rates}

Whether the NS is accreting from the ISM and at what rate depends on the local density of the ISM and properties of the NS (e.g., its mass, radius and magnetosphere; see, e.g. \citealt{1993ApJ...403..690B}). Furthermore, when considering central compact objects (CCOs), the NS is within a supernova remnant, for which the distribution and composition differ from the ISM (see, e.g. \citealt{2012A&ARv..20...49V,2018SSRv..214...44L}).

Following classical accretion physics (see, e.g. \cite{1983bhwd.book.....S}), the accretion rate onto the NS surface can be described using the Bondi formula

\begin{equation}
\label{eq:bondi}
\dot{M} = \frac{4 \pi \lambda_a G^2 M^2 \rho_{\rm ISM}}{\left( v^2 + c_s^2 \right)^{3/2}},
\end{equation}

\noindent
where $\rho_{\rm ISM}$ is the local interstellar mass density, $v$ is relative speed between the NS and ISM, and $c_s$ is the speed of sound. The parameter $\lambda_a$ can vary between 0.25 and 1 \citep{1971MNRAS.154..141H}. Here we take $c_s$ = 10 km s$^{-1}$ and $\lambda_a$ = 1.

For simplicity, we consider three constant values of mass accretion rate for our models, i.e., we do not account for potential short time scale variability in the accretion rate. In the present work, we consider only the effect of a light element column density on the resulting cooling curve and ignore the effect of heat generated by nuclear reactions (see Section \ref{sec:discussion}). We can obtain the highest accretion rate considered here 
($\dot{M}_{\rm acc}=10^{-15}\mbox{ $\Msun$ yr$^{-1}$}$)
by using Equation \ref{eq:bondi} for an isolated $1.4\,\Msun$ NS with
$v=20\mbox{ km s$^{-1}$}$ and ISM number density $n=1\mbox{ cm$^{-3}$}$.
Lower accretion rate values 
($10^{-18}$ and $10^{-21}\mbox{ $\Msun$ yr$^{-1}$}$) are possible for higher
velocities (e.g., 200 or $1000\mbox{ km s$^{-1}$}$) and/or lower ISM densities. 
These low accretion rates are further motivated by possible suppression of the Bondi accretion rate such that NSs in this case accrete at lower rates (see, e.g., \citealt{1975A&A....39..185I,2003ApJ...594..936P,2005ApJ...618..757K}).

\subsection{Cooling curves}

\begin{figure}
	\includegraphics[width=\columnwidth]{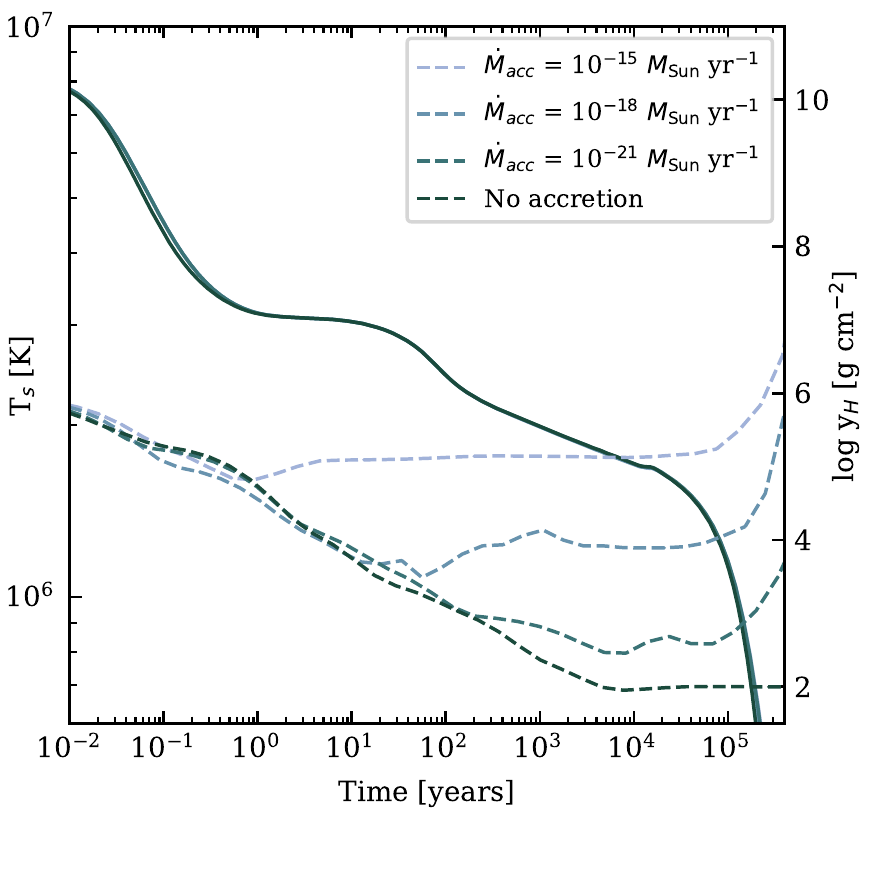} 
	\vspace{-20pt}
	\caption{Cooling curves with initial hydrogen column density y$_{\text{H}}$~=~10$^{6}$~g~cm$^{-2}$ (solid curves) for different mass accretion rates. Dashed lines show the corresponding evolution of envelope composition (right axis). }
	\label{fig:coolingcompositionH}
\end{figure}

\begin{figure}
	\includegraphics[width=\columnwidth]{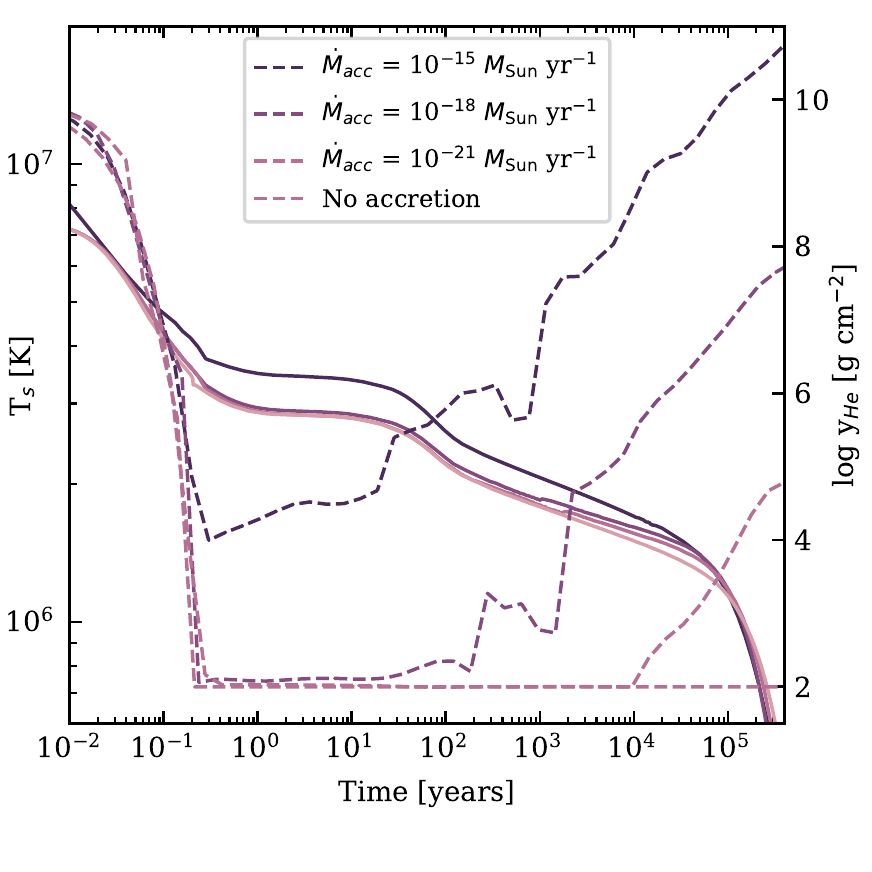}
	\vspace{-20pt}
	\caption{Cooling curves with initial helium column density y$_{\text{He}}$~=~10$^{10}$~g~cm$^{-2}$ (solid curves) for different mass accretion rates. Dashed lines show the corresponding evolution of envelope composition (right axis). }
	\label{fig:coolingcomposition}
\end{figure}

For H-He envelopes, we show the cooling curves and hydrogen column evolution in Figure \ref{fig:coolingcompositionH} for an initial hydrogen column density of $y_{\rm H}$ = $10^{6}$ g cm$^{-2}$; note that the column evolution curves here and in Figures~\ref{fig:coolingcomposition} and \ref{fig:coolingcompositionHC} are smoothed (averaged over short time intervals) in order to remove short timescale fluctuations due to the finite time steps of the numerical simulations. The first conclusion that is clear from this plot is that the cooling curve is almost unaffected by changes in hydrogen column density, as we already found in Section \ref{sec:tbts}. Nevertheless, it is interesting to see how the hydrogen column changes over time for varying ISM accretion rates. Note that for both the H-He and He-C models, we have artificially set the lower limit for the light element column at 100 g cm$^{-2}$. Changes in the light element column below this limit will not affect the cooling curve, as they occur at low densities outside of the sensitivity strip. If DNB is the only mechanism for depleting surface hydrogen, the column density initially drops sharply due to high temperatures, until hydrogen reaction rates become limited by the amount of material available in the burning region. In other words, when the temperature has dropped significantly, the burning is limited by the time it takes for hydrogen to diffuse to the burning depth. Thus for most of such a NS's life (when $\Ts\lesssim 10^{6}\mbox{ K}$), DNB in the NS envelope is not limited by the nuclear reaction rate, but by the diffusion time. For the two scenarios with higher accretion rates we find that, after a period of accreting and burning at similar rates ($\gtrsim$ 10$^5$ years), the accretion rate dominates, and the hydrogen column grows.

In Figure \ref{fig:coolingcomposition}, we show cooling curves and helium column density evolution for an He-C envelope with an initial helium column density of $y_{\rm He}$ = $10^{10}$ g cm$^{-2}$ for different accretion rates. In all cases where $\dot{M}_{\rm acc}$ is less than the maximum accretion rate, more than 99$\%$ of the initial helium is consumed at the hot early stages and the helium column density drops to $y_{\rm He}$~$<$~2~$\times$~10$^{6}$~g~cm$^{-2}$ within 1 year. During this time, helium is more rapidly burned than it is accreted. Depending on the accretion rate, the temperature has dropped enough at $t\sim 1 - 100$ years for the burning rate to become comparable to the accretion rate. After this time (when $\dot{y}_{\rm acc} > \zeta_{\rm DNB}$), the helium column gradually increases as matter accumulates from the ISM, and the temperature is too low for DNB to sufficiently counteract. The scenario with no accretion results in an effectively constant low column density after the initial helium column is depleted. The resulting cooling curve at times $>$ 1 year is the same as a cooling curve with no helium column.The cooling curves for the scenarios with larger accretion rates show short timescale variations, corresponding to variations in helium column. Although the short timescale changes in column density can be large, the effect on the surface temperature is small ($\sim$ a few percent), as the $\Ts-\Tb$ relations are only sensitive to changes in the column density that affect the composition in the sensitivity strip.

\subsection{Implications for photospheric composition}
\label{sec:photosphere}

In the discussion above, we focus on the effect of DNB on the composition of the envelope and the resulting temperature relations between the surface and the interior. We now consider the implications of the envelope evolution for the composition at the photosphere, which is uncertain for isolated NSs and is important for interpreting observations of NS surface radiation.

In Figures \ref{fig:coolingcompositionH} and \ref{fig:coolingcomposition}, the (artificially set) minimum light element column of  $y_{\text{light}}$ = 100 g cm$^{-2}$ is large enough such that the photosphere, and corresponding spectrum, will always be dominated by hydrogen and helium, respectively. Due to this large lower limit, these figures do not directly address whether or not heavier elements can be present in the photosphere. Nevertheless Figure \ref{fig:coolingcompositionH} shows that, even at the low accretion rate of 10$^{-21} \Msun$ yr$^{-1}$, the chosen minimum column is never reached, as the hydrogen column plateaus and increases after a few thousand years. Thus for a H-He envelope, the photosphere will be dominated by hydrogen even for very low accretion rates. On the other hand, Figure \ref{fig:coolingcomposition} shows that, for accretion rates $< 10^{-15} \Msun$ yr$^{-1}$, there will be a period after $\sim$1 month, where the spectrum of the He-C envelope can be dominated by carbon.

\begin{figure}
	\includegraphics[width=\columnwidth]{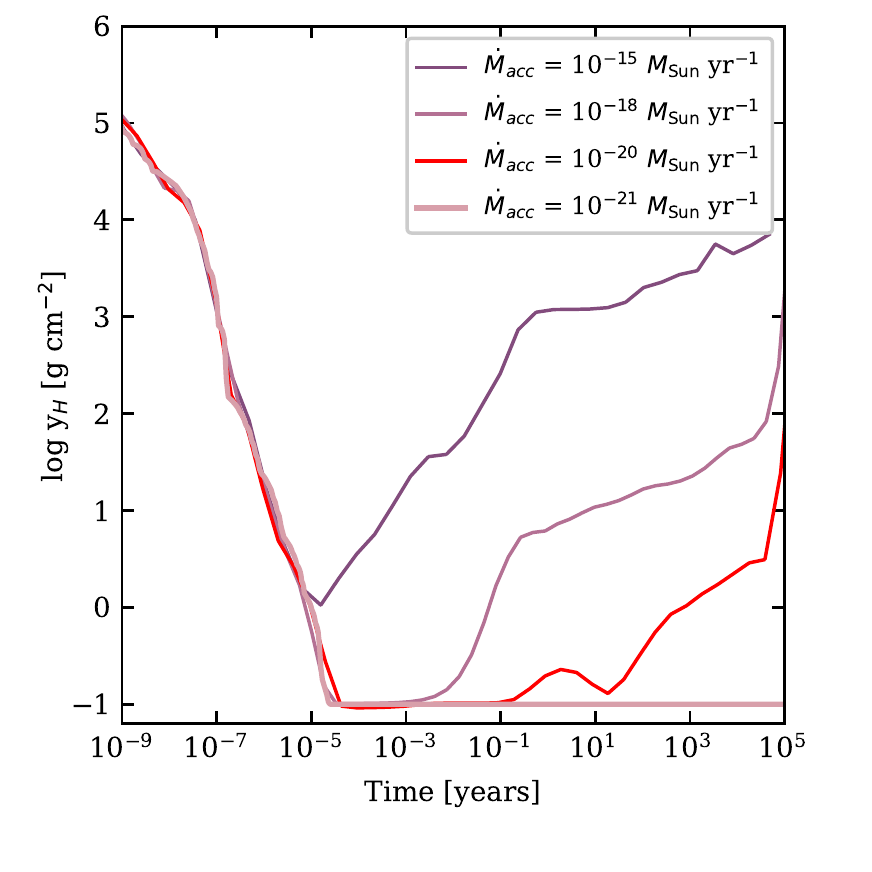}
	\vspace{-20pt}
	\caption{Evolution of the hydrogen column with an initial column density y$_{\rm H}$~=~10$^{6}$~g~cm$^{-2}$ for different mass accretion rates. }
	\label{fig:coolingcompositionHC}
\end{figure}

As accreted material from the ISM consists mainly of hydrogen, and thus accretion will mostly make the photosphere (and envelope) more hydrogen rich, it is interesting to consider the evolution of a H-C envelope. Therefore, we compute a H-C envelope model with $\rho_b$ = 10$^{10}$ g cm$^{-3}$ and $g_s$ = 2.43 $\times$ 10$^{14}$ cm s$^{-2}$, for which the possible nuclear reactions are proton-proton captures and proton captures onto carbon (see \citealt{2004ApJ...605..830C}). In Figure \ref{fig:coolingcompositionHC} we show the column evolution for an initial hydrogen column of $y_{\text{H}}$ = 10$^{6}$ g cm$^{-2}$ using a lower limit of 0.1 g cm$^{-2}$. Note that this lower limit corresponds to a hydrogen fraction at the photosphere of $\sim$5$\%$ and thus a carbon dominated spectrum. Only slightly larger hydrogen columns are enough to result in a hydrogen spectrum, as a hydrogen column of $\sim 1.5$ g cm$^{-2}$ corresponds to a hydrogen fraction of $\sim$90$\%$ at the photosphere. Figure \ref{fig:coolingcompositionHC} shows that due to the highly efficient capture onto carbon, the initial hydrogen column is rapidly depleted within an hour for accretion rates $< 10^{-15}  \Msun$ yr$^{-1}$. For these accretion rates, a period can exist when the spectrum is that of a carbon atmosphere. An accretion rate of $10^{-20}  \Msun$ yr$^{-1}$ leads to an optically thick hydrogen atmosphere after $\sim$1000 year. For higher accretion rates, the hydrogen column will effectively always be large enough to maintain a hydrogen atmosphere. On the other hand, accretion rates $< 10^{-21}  \Msun$ yr$^{-1}$ are too low to build up an optically thick hydrogen column and will result in a carbon atmosphere after the initial hydrogen column is consumed. These results are relevant for the understanding of the neutron star in Cas A, of which the spectra can be well described by a carbon atmosphere \citep{2009Natur.462...71H}, as well as other NSs for which a carbon atmosphere is suggested (see Section \ref{sec:discussion}). We will consider temperature relations for H-C envelopes and their application to these neutron stars in future work.

\section{Application to Cassiopeia A}
\label{sec:casa}
\subsection{\Chandra\ data analysis} \label{sec:data}

We process all \Chandra\ ACIS-S GRADED data using Chandra Interactive Analysis
of Observations (\textsc{CIAO}) 4.10 and Chandra Calibration Database
(\textsc{CALDB}) 4.8.1, following the procedure described in \cite{2010ApJ...719L.167H}. 
Spectra are extracted using \textsc{specextract} from a 4 pixel circular region
centered on the source, while background is taken from a 5$-$8 pixel annulus
centered on the source.
Observations taken close in time (see Table~\ref{tab:casa}) are merged using
\textsc{combine\_spectra} and \textsc{dmgroup} with a minimum of 200 counts
per energy bin.

Spectra are fit with \textsc{Xspec} 12.10.0c \citep{1996ASPC..101...17A} and
a four component model consisting of
\textsc{pileup}, \textsc{tbabs}, \textsc{spexpcut}, and \textsc{nsx}.
For \textsc{pileup} \citep{2001ApJ...562..575D}, we set the grade migration
parameter to 0.27 for observations through 2004, which have a frame time of
3.24~s, and to 0.24 for observations starting from 2007, which have
a frame time of 3.04~s; the maximum number of photons is set to 5, and the
point-spread-function fraction is set to 0.95
(see \citealt{2010ApJ...719L.167H}, for details).
To model photoelectric absorption, we use \textsc{tbabs} with abundances from
\citet{2000ApJ...542..914W}.
To model interstellar dust scattering, we use \textsc{spexpcut} with exponent
index $\alpha=-2$ and characteristic energy
$E_{\rm cut}(\mbox{keV})=[0.49\NH(10^{22}\mbox{ cm$^{-2}$})]^{1/2}$
\citep{2003AN....324...73P}, where $\NH$ is the absorption column determined
from \textsc{tbabs}.
For \textsc{nsx} \citep{2009Natur.462...71H}, we assume a partially ionized
carbon atmosphere and fix the distance to 3.4~kpc \citep{1995ApJ...440..706R}
and normalization to 1.

\subsection{Spectral fits} \label{sec:spectralfit}
To determine the neutron star mass $M$ and radius $R$ for use in spectral
fitting, we first allow fit parameters $M$, $R$, $\NH$, and $\Ts$ to vary
but tie them between observations, so that each has the same value across
all observations.
The resulting fit produces best-fit values of
$M=1.44\pm0.15\,\Msun$, $R=13.78\pm0.36\mbox{ km}$,
$\NH=(1.623\pm0.030)\times 10^{22}\mbox{ cm$^{-2}$}$,
and $\Ts=(1.728\pm0.030)\times 10^6\mbox{ K}$,
with $\chi^2=1646$ for 1458 degrees of freedom (dof).
If we instead allow $\Ts$ to have different values between observations,
then the best-fit values are
$M=1.65\pm0.16\,\Msun$, $R=12.94\pm0.34\mbox{ km}$,
and $\NH=(1.674\pm0.031)\times 10^{22}\mbox{ cm$^{-2}$}$,
with $\chi^2/\mbox{dof}=1522/1446$.
If we allow both $\NH$ and $\Ts$ to have different values between
observations, then the best-fit values are
$M=1.46\pm0.16\,\Msun$ and $R=13.69\pm0.36\mbox{ km}$,
with $\chi^2/\mbox{dof}=1497/1434$.
A f-test between the first and second fits ($\Delta\chi^2=124$) and between
the second and third fits ($\Delta\chi^2=25$) yield probabilities of
$9.8\times 10^{-19}$ and 0.024, respectively, which suggest that allowing $\Ts$, and possibly $\NH$, to change between observations is warranted.
Therefore in the spectral fits performed below, we untie $\Ts$ and fix the
neutron star mass and radius to $M=1.65\,\Msun$ and $R=12.9\mbox{ km}$,
respectively.

\begin{figure}
	\includegraphics[width=0.9\columnwidth]{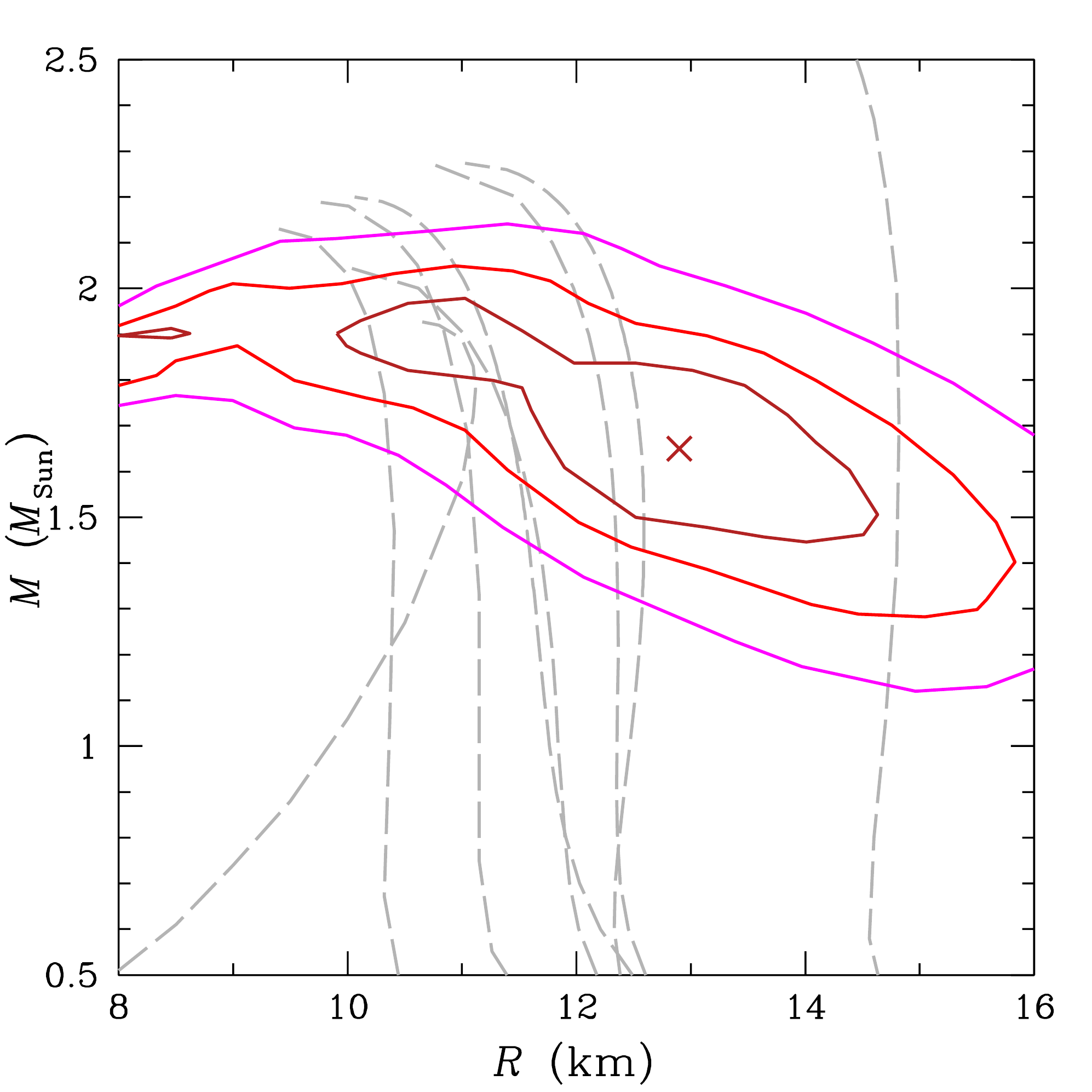}
	\caption{
		Neutron star mass $M$ and radius $R$.
		Solid lines are the 1$\sigma$, 90\%, and 99\% confidence contours on $M$
		and $R$ obtained from fitting \Chandra\ spectra of the neutron star in the
		Cassiopeia~A supernova remnant, assuming $\NH$ does not change between
		observations.
		Cross marks the best-fit $(M,R)=(1.65\,\Msun,12.9\mbox{ km})$.
		Dashed lines indicate mass-radius relation obtained from a sample of
		theoretical nuclear equations of state
		(see \citealt{2001A&A...380..151D,2001ApJ...550..426L,2013A&A...560A..48P,2014PhRvC..89d8801B},
		and references therein).
	}
	\label{fig:mr}
\end{figure}

We show the $M$-$R$ confidence contour in Fig.~\ref{fig:mr}, assuming a distance of 3.4~kpc. The best-fit values from our previous work
\citep{2010ApJ...719L.167H} and used in subsequent works
\citep{2011MNRAS.412L.108S,2013ApJ...777...22E,2015PhRvC..91a5806H}
are $M=1.65\,\Msun$ and $R=10.3\mbox{ km}$. It is important to note that, while the $M$-$R$ constraints we obtain are somewhat constraining, there are caveats. First, there are calibration issues with the data, which are discussed in Section \ref{sec:discussion} (see also \citealt{2013ApJ...779..186P}). Second, the magnetic field strength of Cas~A is assumed here to be low enough as to not affect its atmospheric emission. This may not necessarily be the case, as three other CCOs have surface magnetic fields $\sim 10^{10}-10^{11}\mbox{ G}$ \citep{2013ApJ...765...58G,2013IAUS..291..101H,2017JPhCS.932a2006D}, which is somewhat larger than accommodated for in our atmosphere model (c.f. \citealt{2014A&A...572A..69P}). Finally, it is possible that the NS surface is not uniform in temperature, and hot spots could dominate the observed emission (see, e.g., \citealt{2010ApJ...724.1316G,2014ApJ...790...94B}), although pulsations from this emission are not detected in Cas~A \citep{2002ApJ...566.1039M,2010ApJ...709..436H}.

\begin{figure}
	\includegraphics[width=0.9\columnwidth]{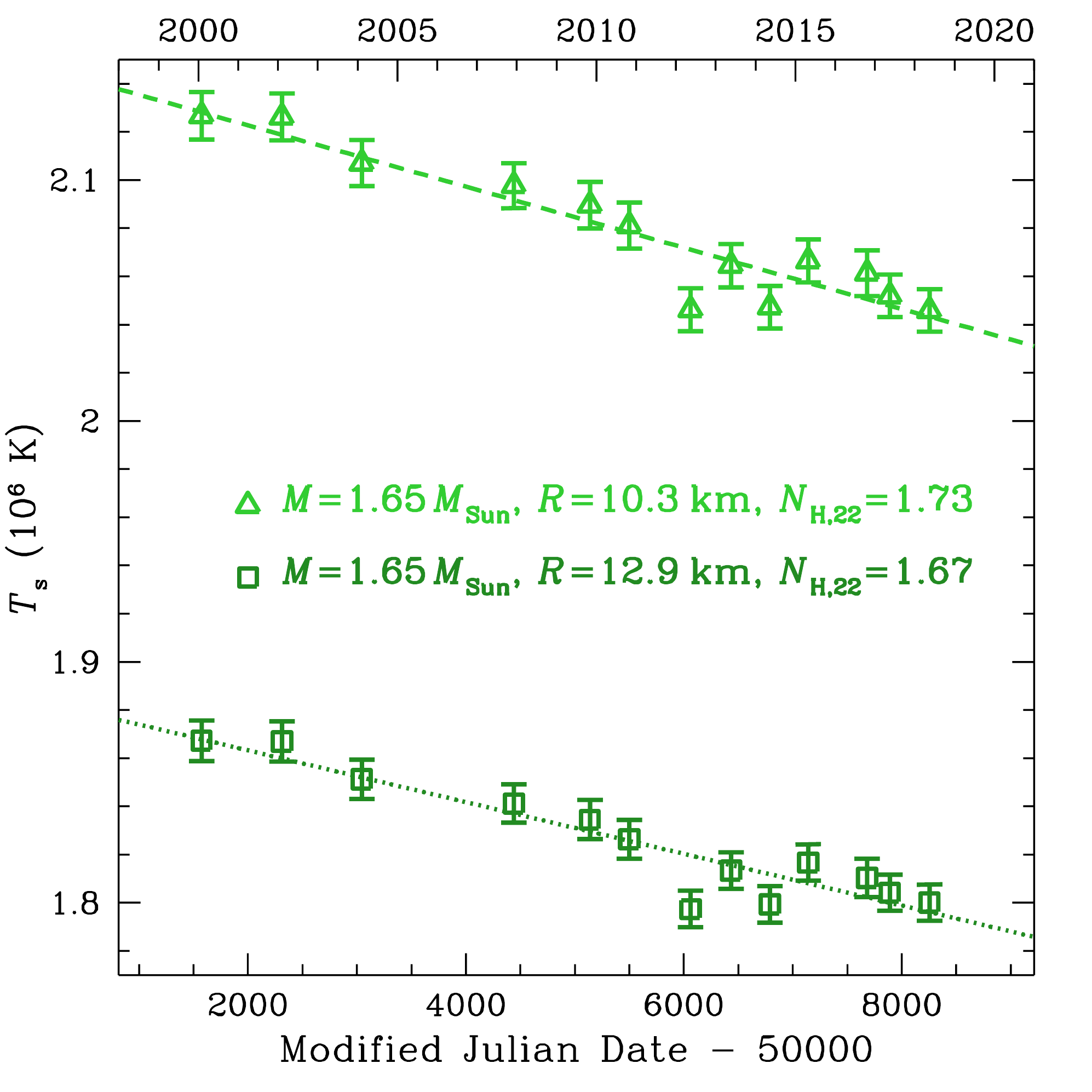}
	\caption{
		Surface temperature of the neutron star in the Cassiopeia~A supernova remnant as
		measured from \Chandra\ ACIS-S GRADED spectra over the past 18 years.
		Triangles indicate $\Ts$ obtained using the best-fit neutron star mass
		$M=1.65\,\Msun$ and radius $R=10.3\mbox{ km}$ from \citet{2010ApJ...719L.167H} and
		absorption column $\NHtwo\equiv\NH/10^{22}\mbox{ cm$^{-2}$}=1.73$ from
		\citet{2013ApJ...777...22E},
		while squares indicate $\Ts$ obtained using updated best-fit $M=1.65\,\Msun$,
		$R=12.9\mbox{ km}$, and $\NHtwo=1.67$ (see text).
		Error bars are $1\sigma$.
		Dashed and dotted lines show linear fits to each set of $\Ts$.
	}
	\label{fig:tempnhfreeze}
\end{figure}

\begin{figure}
	\includegraphics[width=0.9\columnwidth]{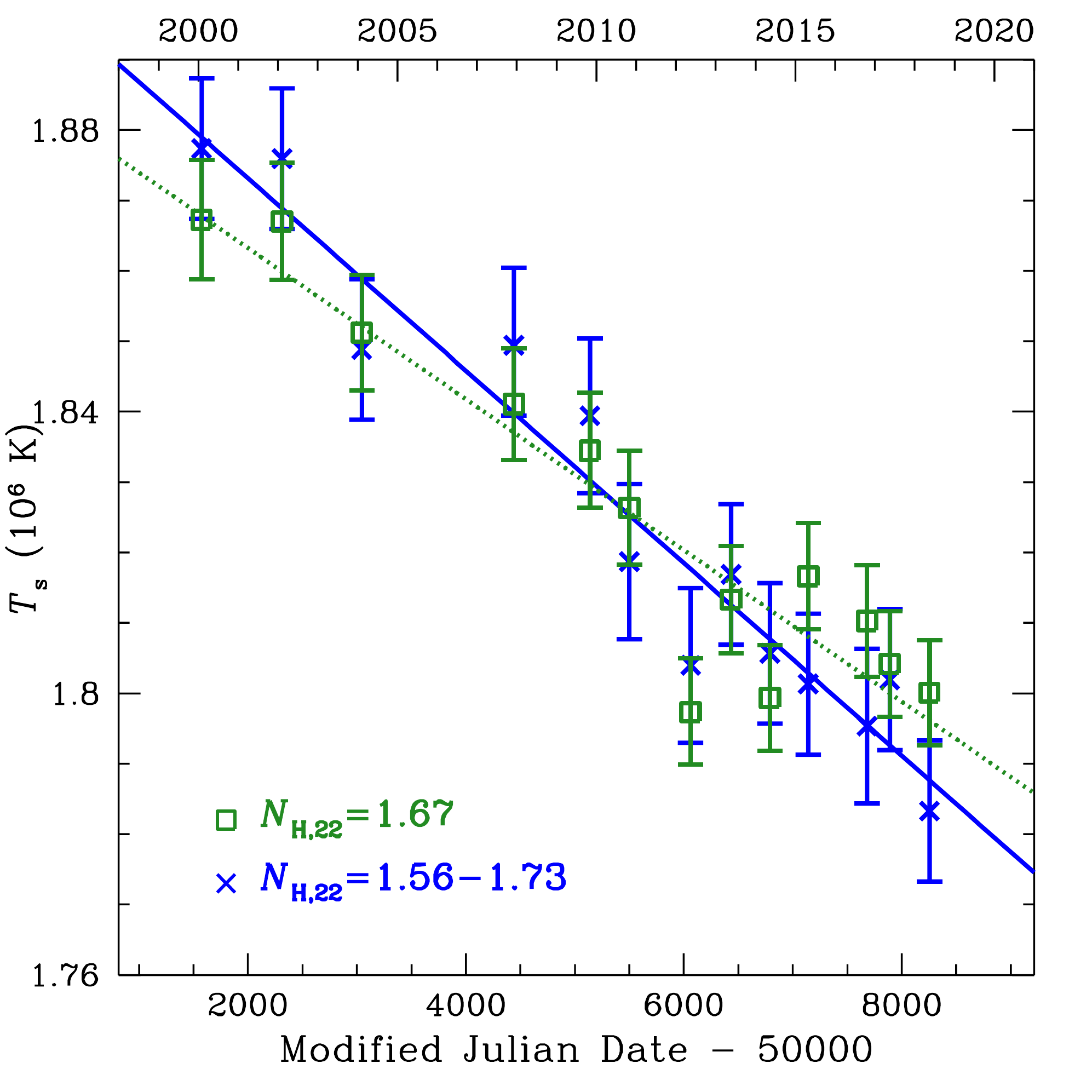}
	\caption{
		Surface temperature of the neutron star in the Cassiopeia~A supernova remnant as
		measured from \Chandra\ ACIS-S GRADED spectra over the past 18 years.
		Data points indicate $\Ts$ obtained using best-fit neutron star mass
		$M=1.65\,\Msun$ and radius $R=12.9\mbox{ km}$ and a constant absorption column
		$\NHtwo\equiv\NH/10^{22}\mbox{ cm$^{-2}$}=1.67$ (squares) and changing $\NH$
		(crosses; see Fig.~\ref{fig:nh}).
		Error bars are $1\sigma$.
		Dotted and solid lines show linear fits to each set of $\Ts$.
	}
	\label{fig:temp}
\end{figure}

\begin{figure}
	\includegraphics[width=0.9\columnwidth]{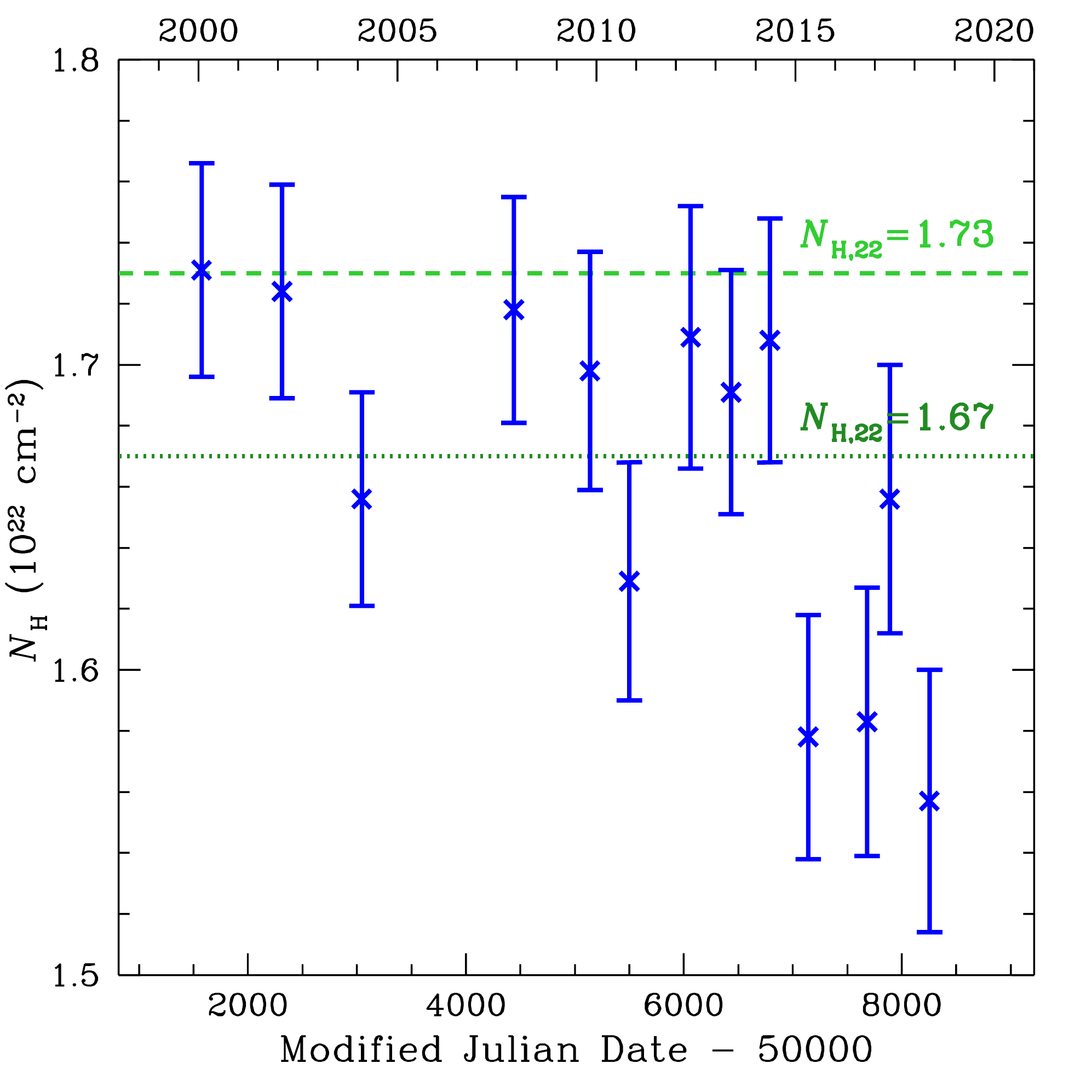}
	\caption{
		Absorption column $\NH$ as measured from a fit to \Chandra\ spectra of the
		Cassiopeia~A neutron star.
		Data points indicate $\NH$ using best-fit neutron star mass $M=1.65\,\Msun$
		and radius $R=12.9\mbox{ km}$ and changing $\Ts$,
		while dotted and dashed lines indicate constant
		$\NHtwo\equiv\NH/10^{22}\mbox{ cm$^{-2}$}=1.67$ and 1.73, respectively, used
		in other spectral fits (see Figs.~\ref{fig:tempnhfreeze} and \ref{fig:temp}).
		Error bars are $1\sigma$.
	}
	\label{fig:nh}
\end{figure}

Next, we perform two fits of all spectra with the same set of models but
fixing the parameters $M=1.65\,\Msun$ and $R=12.9\mbox{ km}$.
The first fit allows $\Ts$ to vary between observations but holds the
absorption parameter fixed at $\NH=1.67\times 10^{22}\mbox{ cm$^{-2}$}$.
This fit yields $\chi^2/\mbox{dof}=1522/1449$, and the temperature evolution
is shown in Fig.~\ref{fig:tempnhfreeze} and Table~\ref{tab:casa}.
The second fit allows both $\Ts$ and $\NH$ to change between observations.
This fit yields $\chi^2/\mbox{dof}=1498/1436$, and temperature and
absorption column evolutions are shown in Figs.~\ref{fig:temp} and \ref{fig:nh},
respectively (see also Table~\ref{tab:casa}).
A f-test between these two fits ($\Delta\chi^2=24$) yields a probability of
0.043, so allowing $\NH$ to vary is possibly justified (see Section \ref{sec:discussion}).
For comparison to our previously published results, we also perform a fit using
fixed
$(M,R,\NH)=(1.65\,\Msun,10.3\mbox{ km},1.73\times 10^{22}\mbox{ cm$^{-2}$})$.
Results are shown in Fig.~\ref{fig:tempnhfreeze} and Table~\ref{tab:casa} and
have $\chi^2/\mbox{dof}=1562/1449$.
The approximately constant offset in temperatures (e.g., $\sim 2.13/1.87=1.14$)
between the two fixed-$\NH$ fit results plotted in Fig.~\ref{fig:tempnhfreeze}
is primarily due to the difference in assumed $R$ and its anti-correlation with
$\Ts$, i.e., for constant luminosity, $\Ts\propto R^{-1/2}$, which yields
$(12.9\mbox{ km}/10.3\mbox{ km})^{1/2}=1.12$, as well as a difference in
gravitational redshift.

\begin{table*}
	\caption{
		Surface temperature $\Tssix$ ($10^{6}\mbox{ K}$), absorbed 0.5--7~keV flux $F_{\rm -13}^{\rm abs}$
		($10^{-13}\mbox{ erg cm$^{-2}$ s$^{-1}$}$), and absorption column
		$\NHtwo$ ($10^{22}\mbox{ cm$^{-2}$}$) determined from model fits to
		\Chandra\ ACIS-S GRADED spectra of the neutron star in the Cassiopeia~A
		supernova remnant.
		Three spectral fit results are shown, two with constant $\NH$ and one with
		changing $\NH$.
		Each set of 13 temperatures and fluxes are fit to a linear decline, with
		decline rate and fit statistic as shown.
		For merged ObsIDs, the MJD listed is that of the first ObsID.
		Number in parentheses is $1\sigma$ uncertainty in last digit.
	}
	\label{tab:casa}
	\begin{tabular}{rlcccccccc}
		\hline
		& & $\NHtwo$ & \multicolumn{2}{c}{1.73} & \multicolumn{2}{c}{1.67} & \multicolumn{3}{c}{see below} \\
		& & $M$ ($\Msun$) & \multicolumn{2}{c}{1.65} & \multicolumn{2}{c}{1.65} & \multicolumn{3}{c}{1.65} \\
		& & $R$ (km) & \multicolumn{2}{c}{10.3} & \multicolumn{2}{c}{12.9} & \multicolumn{3}{c}{12.9} \\
		& & $\chi^2$/dof & \multicolumn{2}{c}{1562/1449} & \multicolumn{2}{c}{1522/1449} & \multicolumn{3}{c}{1498/1436} \\ \\
		ObsID & Date & MJD & $\Tssix$ & $F_{\rm -13}^{\rm abs}$ & $\Tssix$ & $F_{\rm -13}^{\rm abs}$ & $\Tssix$ & $F_{\rm -13}^{\rm abs}$ & $\NHtwo$ \\
		\hline
		114 & 2000 Jan 30 & 51573.4 & 2.127(10) & 7.4(2) & 1.867(8) & 7.3(1) & 1.877(10) & 7.3(2) & 1.73(3) \\
		1952 & 2002 Feb 6 & 52311.3 & 2.126(10) & 7.4(2) & 1.867(8) & 7.3(1) & 1.876(10) & 7.4(2) & 1.72(3) \\
		5196 & 2004 Feb 8 & 53043.7 & 2.107(10) & 7.1(1) & 1.851(8) & 7.0(2) & 1.849(10) & 7.0(2) & 1.66(3) \\
		9117/9773 & 2007 Dec 5/8 & 54439.9 & 2.098(9) & 7.0(2) & 1.841(8) & 6.9(2) & 1.849(11) & 6.9(2) & 1.72(4) \\
		10935/12020 & 2009 Nov 2/3 & 55137.9 & 2.090(10) & 6.9(2) & 1.834(8) & 6.8(2) & 1.839(11) & 6.8(1) & 1.70(4) \\
		10936/13177 & 2010 Oct 31/Nov 2 & 55500.2 & 2.081(10) & 6.8(2) & 1.826(8) & 6.7(2) & 1.819(11) & 6.6(1) & 1.63(4) \\
		14229 & 2012 May 15 & 56062.4 & 2.046(9) & 6.4(2) & 1.797(8) & 6.3(1) & 1.804(11) & 6.4(1) & 1.71(4) \\
		14480 & 2013 May 20 & 56432.6 & 2.064(9) & 6.6(2) & 1.813(8) & 6.5(1) & 1.817(10) & 6.5(1) & 1.69(4) \\
		14481 & 2014 May 12 & 56789.1 & 2.047(9) & 6.3(2) & 1.799(7) & 6.2(1) & 1.806(10) & 6.3(1) & 1.71(4) \\
		14482 & 2015 Apr 30 & 57142.5 & 2.066(9) & 6.7(1) & 1.817(8) & 6.6(1) & 1.801(10) & 6.6(1) & 1.58(4) \\
		19903/18344 & 2016 Oct 20/21 & 57681.2 & 2.061(9) & 6.6(2) & 1.810(8) & 6.5(2) & 1.795(11) & 6.5(1) & 1.58(4) \\
		19604 & 2017 May 16 & 57889.7 & 2.052(9) & 6.5(2) & 1.804(7) & 6.4(2) & 1.802(10) & 6.4(2) & 1.66(4) \\
		19605 & 2018 May 15 & 58253.7 & 2.046(9) & 6.4(2) & 1.800(7) & 6.4(1) & 1.783(10) & 6.4(1) & 1.56(4) \\
		\hline
		& & 10-year & & & & & & & \\
		& & decline rate & $2.2\pm0.2\%$ & $7.5\pm1.0\%$ & $2.1\pm0.2\%$ & $7.3\pm0.9\%$ & $2.7\pm0.3\%$ & $7.7\pm0.9\%$ & \\
		& & $\chi^2$/dof & 14.8/11 & 11.0/11 & 15.7/11 & 15.8/11 & 6.2/11 & 10.9/11 & \\
		\hline
	\end{tabular}
\end{table*}

Finally, we perform a linear fit to each set of declining temperatures $\Ts$.
For
$(M,R,\NH)=(1.65\,\Msun,10.3\mbox{ km},1.73\times 10^{22}\mbox{ cm$^{-2}$})$,
the slope is $-13\pm1\mbox{ K d$^{-1}$}$ ($\approx -2.2\pm0.2\%$ per
10 years) with $\chi^2/\mbox{dof}=14.8/11$.
For
$(M,R,\NH)=(1.65\,\Msun,12.9\mbox{ km},1.67\times 10^{22}\mbox{ cm$^{-2}$})$,
the slope is $-11\pm1\mbox{ K d$^{-1}$}$ ($\approx -2.1\pm0.2\%$ per
10 years) with $\chi^2/\mbox{dof}=15.7/11$.
For $(M,R)=(1.65\,\Msun,12.9\mbox{ km})$ and changing $\NH$,
the slope is $-14\pm1\mbox{ K d$^{-1}$}$ ($\approx -2.7\pm0.3\%$ per
10 years) with $\chi^2/\mbox{dof}=6.2/11$. We discuss uncertainties beyond statistical ones in our measured temperature decline rates and compare to rates measured in previous works in Section \ref{sec:discussion}.

\subsection{Cas A cooling curve}
\label{sec:CasA_cooling}

The observed temperature drop has been interpreted as a result of neutron superfluidity and proton superconductivity in Cas A (\citealt{2011PhRvL.106h1101P,2011MNRAS.412L.108S,2011MNRAS.411.1977Y,2015PhRvC..91a5806H}, see Section \ref{sec:intro} for references on alternative interpretations). With the addition of 4 new data points presented above, we now have a set of 13 temperatures spanning over 18 years. We apply the cooling model described in Section \ref{sec:coolmodel} to the full set of data. We use the temperature data corresponding to the spectral fits with $(M,R,\NH)=(1.65\,\Msun,12.9\mbox{ km},1.67\times 10^{22}\mbox{ cm$^{-2}$})$. The cooling model makes use of a time-variable He-C envelope model with an initial He column of 10$^6$ g cm$^{-2}$ and assumes no accretion from the ISM. As we found in Section \ref{sec:timedependent}, the cooling curve is not sensitive to the initial helium column density, as practically all helium will be depleted within 10 years. Therefore, the cooling curve at the time of the Cas A observations (at age $\approx 320-340\mbox{ yr}$) is the same as a cooling curve using a pure carbon envelope.

In order to describe the rapid temperature decline, we follow the previous works (e.g., \citealt{2011PhRvL.106h1101P,2011MNRAS.412L.108S}) by invoking superfluidity of neutrons in the inner crust through pairing in the $^1S_0$ state and in the core through pairing in the $^3P_2$ state. Within the minimal cooling paradigm \citep{2004A&A...423.1063G,2004ApJS..155..623P}, this means that stronger neutrino emission is only allowed as a result of neutron or proton pairing. The exact properties, such as the critical temperature $\Tc$, of neutron $^3P_2$ superfluidity in the core are uncertain and can be constrained by comparing the cooling curve to the observed temperatures \citep{2011PhRvL.106h1101P,2011MNRAS.412L.108S,2015PhRvC..91a5806H}. In our model here, we use superfluidity and superconductivity properties similar to those in \cite{2011PhRvL.106h1101P}; specifically, we use a neutron $^3P_2$ superfluid model with a maximum critical temperature $\Tc\sim 5.9 \times 10^8\mbox{ K}$ for $M=1.7\,\Msun$ (with the APR EOS, resulting in $R$ = 11.4 km) and the `CCDK' proton $^1S_0$ superconducting model described in \cite{2004ApJS..155..623P}. Note that our considered theoretical mass and radius using the APR EOS (1.7 Msun and 11.4 km) is within the 90 percent confidence contour of mass and radius obtained from the spectral analysis (see Figure \ref{fig:mr}). The aim of our calculations here is not to find a best fitting model but to show that models fitting the new data are in line with previous results for superfluid properties and can also be explained using the envelope $\Tb-\Ts$ relations that include DNB.

\begin{figure}
	\includegraphics[width=\columnwidth]{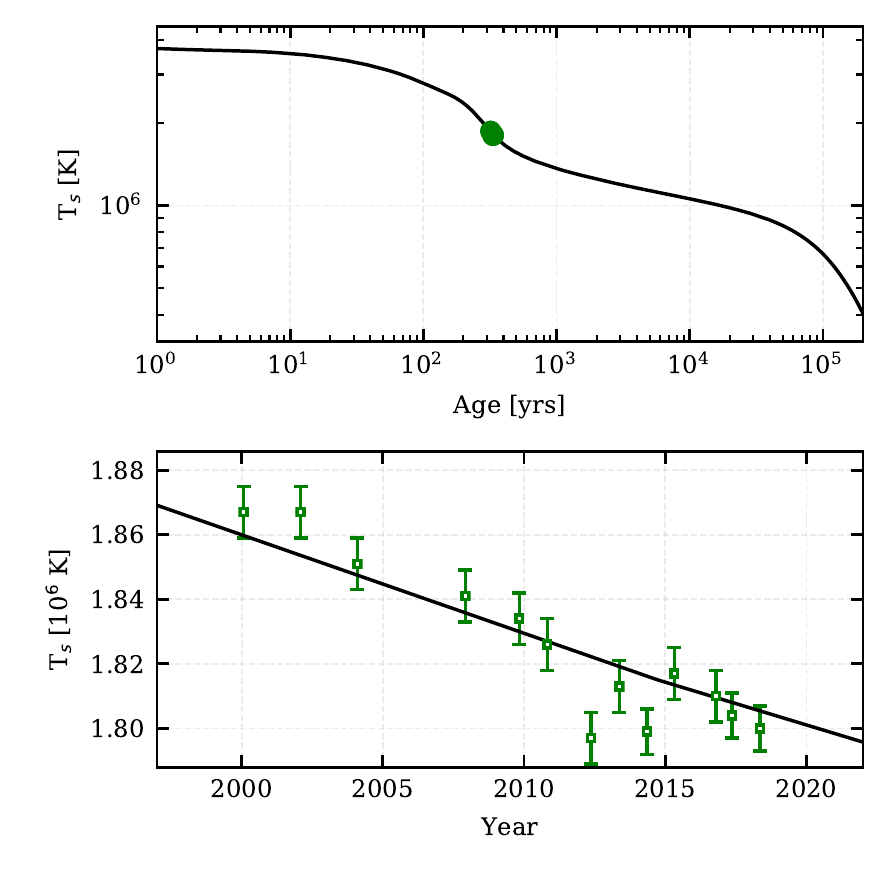} 
	\vspace{-20pt}
	\caption{Example cooling curve for Cas A with $M$ = 1.7 $\Msun$ and $R$ = 11.4 km. Model parameters are similar to those reported in \citet{2011PhRvL.106h1101P}. The rapid temperature decline at $\sim 340$~yr is explained by a neutron $^3P_2$ superfluid in the core with a maximum $\Tc\approx 5.9 \times 10^8\mbox{ K}$. }
	\label{fig:casa_cool}
\end{figure}

To compare the model to the observations, we use the date for the birth of the Cas A NS of $1681\pm19$ \citep{2006ApJ...645..283F}. The resulting cooling curve is shown in Figure \ref{fig:casa_cool} and can adequately describe the observed temperatures. It is important to note that several combinations of superfluid and neutron star properties can describe the observed cooling, so the model presented here is not a unique solution and is merely for illustrative purposes.
   
\subsection{Implications of DNB on Cas A photosphere}
\label{sec:dnbCasA}

In Section 5.5, we show that DNB rapidly consumes any initial hydrogen from H-C
envelopes for low to moderate accretion rates, thus allowing for an atmosphere
composed of carbon to be visible.  Effectively, hydrogen atmosphere formation
by accretion in the dense post-supernova environment is delayed by DNB to a
later, lower density epoch.  For the relatively young Cas A NS, this is likely
the case.  In order for Cas A to have a carbon atmosphere, it must accrete at
an average rate lower than $10^{-20}\mbox{ $\Msun$ yr$^{-1}$}$; for larger
accretion rates, DNB cannot consume hydrogen fast enough to prevent formation
of a hydrogen atmosphere.  While DNB can explain a carbon atmosphere, it is
unlikely to be responsible for the observed temperature variations in Cas A,
as the light element burning rate is negligible when the atmosphere is
dominated by carbon.

\section{Discussion}
\label{sec:discussion}

In this present work, we consider neutron star envelopes consisting of either H-He or He-C and including the physics of diffusive nuclear burning, in order to investigate how DNB impacts the thermal evolution of NSs. DNB is relevant for the thermal evolution in both a static and time-dependent manner: static because DNB affects the interior composition profile of the envelope and thus can alter the $\Ts-\Tb$ relations, and time-dependent because it can alter the composition of the envelope over time. In Section \ref{sec:burning}, we calculate the burning rates and find that DNB can alter the composition on short time-scales ($\lesssim 10-100$ yr). For H-He, the hydrogen burning rate is strongly coupled to the amount of hydrogen available (characterised by its column density) whereas for the He-C envelope the burning rate is highly sensitive to the temperature and is only significant for $\Ts > 10^6$ K.

In Section \ref{sec:tbts}, we obtained static analytic relations for the surface temperature $\Ts$ and the temperature at the bottom of the envelope $\Tb$. These temperature relations (in Appendix \ref{sec:appendix_tbts}) can be used to model the thermal evolution of isolated and accreting neutron stars. We find the following main differences in the $\Ts-\Tb$ relations due to DNB:

\begin{itemize}
	\item {For a H-He envelope, we find that when DNB is taken into account, there is a clear upper limit on the size of the hydrogen column. Larger hydrogen columns ($y_H \gtrsim 10^7$ g cm$^{-2}$) are not thermally stable, as hydrogen is consumed rapidly when DNB is nuclear-limited. For temperatures above $\Ts \sim$ $10^6$ K, this means the spread in $\Tb$ is smaller than without DNB (see Figure \ref{fig:sensitivity2}).}
	\item {For He-C envelopes, DNB is highly sensitive to the temperature, and DNB makes the envelope more transparent to the heat flux at $\Ts > 2 \times 10^{^6}$ K for the helium column sizes that affect the sensitivity strip (see Figure \ref{fig:sensitivity}).} 
\end{itemize}

We then combine these results in Section \ref{sec:timedependent} and use the DNB burning rates (Section \ref{sec:burning}) and the temperature relations (Section \ref{sec:tbts}) to calculate NS cooling curves with envelope compositions that evolve over time through nuclear burning and accretion. As the H-He temperature relations only span a small range of $\Tb$, the changing envelope composition does not impact the cooling curves significantly. We find that if DNB is the only burning mechanism and no accretion takes place, an initial hydrogen column of $10^6$ g cm$^{-2}$ is depleted after $\sim 5 \times 10^3$ years because the burning rate drops as the hydrogen column decreases. For the He-C envelope all helium will be depleted during the hot early cooling phase (within 1 year) if no accretion takes place. Note that we have only considered illustrative cases and have not included heat generated by the nuclear reactions nor the possibility of sudden accretion events. 

For several CCOs, the spectra could be well described by a carbon atmosphere (see, e.g., \citealt{2016A&A...592L..12K,2017A&A...600A..43S,2018A&A...618A..76D}). We consider the photosphere of a H-C envelope to investigate the possibility of a carbon atmosphere on isolated neutron stars. Low levels of accretion lead to an optically thick hydrogen atmosphere, with a hydrogen column of $1.5$ g cm$^{-2}$ corresponding to a hydrogen abundance in the atmosphere of 90$\%$. The composition evolution of the H-C envelope in Figure \ref{fig:coolingcompositionHC} for accretion rates $\leq $ $10^{-20}  \Msun$ yr$^{-1}$ indicate that it is possible for Cas A, at an age of 340 years, to not yet accrete enough hydrogen to form an optically thick hydrogen atmosphere. For older NSs, the possibility of a carbon atmosphere becomes less likely, as even a little bit of accretion leads to a hydrogen dominated atmosphere.

Note that we have only considered constant accretion rates here, whereas scenarios with changing accretion rates are more realistic for CCOs, as their environment is highly variable (see, e.g., \citealt{2012A&ARv..20...49V,2018SSRv..214...44L}). For example, it is possible that the NS initially accretes at very low rates after the nearby environment has been cleared out by the SN shock. Later on, a reverse shock could introduce a higher density of the surrounding medium, and thus a higher accretion rate onto the NS. Early fallback of material could also lead to varying accretion rates. As the variability in the accretion rate and the composition of the accreted material in SNRs are highly model dependent, we have limited our models to illustrative cases with constant rates and compositions.

In all the envelope models currently used in cooling simulations, it is assumed that the flux generated by nuclear reactions in the envelope is negligible compared to the radiative flux from the neutron star interior. However, we find that the assumption of constant flux through the envelope can be invalid in relevant regions of parameter space. The assumption of negligible heat production in the envelope is certainly incorrect for the hot, initial post-supernova stage, when surface temperatures are $\gg 10^6\mbox{ K}$. However, as the energy is produced at low densities ($\rho < 10^8 - 10^{10}$ g cm$^{-3}$) and the interior is still hot, this heat is radiated from the surface and will not affect cooling at later times. Including nuclear heating at early times will likely lead to an even more rapid consumption of light elements than calculated here. At later times, heat generated by DNB is negligible for H-He envelopes, while heat created by DNB can be important for high temperatures (T $\gtrsim$ 2 $\times$ 10$^6$ K) and large helium columns (y$_{\text{He}} \gtrsim 10^{11}$ g cm$^{-2}$) in the case of He-C envelopes. Interestingly, for H-C envelopes the luminosity generated by DNB can be larger than the interior luminosity for hydrogen columns greater than y$_{\text{H}}$ $> 10^{6}$ g cm$^{-2}$ and surface temperatures higher than $\Ts \gtrsim 10^{6}$ K.  We will examine the effect of heat generated by nuclear reactions in future work. 

In Section~\ref{sec:casa}, we present new observations of the NS in the Cassiopeia~A supernova
remnant and apply our model to this data. We performed spectral fits of 13
\Chandra\ ACIS-S GRADED observations of Cas~A taken over more than 18 years (see Table~\ref{tab:casa}).
For constant interstellar absorption
($\NH\approx 1.7\times 10^{22}\mbox{ cm$^{-2}$}$),
these data indicate that the NS temperature is decreasing at a ten-year rate of
$2.1\pm0.2\%$ ($1\sigma$) and a decreasing ten-year flux rate of $\approx 7\%$.
However, it is known that there is a contaminant building up on the \Chandra\ ACIS-S detector that can affect GRADED mode data if uncorrected \citep{2013ApJ...779..186P,2016SPIE.9905E..44P,2018SPIE10699E..6BP}. \cite{2013ApJ...777...22E} analyse data from different detectors and operating modes and infer a ten-year temperature decline rate of $3.5\pm0.4\%$ for ACIS-S GRADED data and a ten-year decline rate of $1.0\pm0.7\%$ for HRC-S data, suggesting that at least one dataset is affected by systematic errors. \citet{2013ApJ...779..186P} analyse an alternative set of 2 \Chandra\ ACIS-S subarray observations taken over 6 years in FAINT mode, to minimize the effects of pile-up and contaminant; they find a lower ten-year rate of flux decline and no apparent temperature decline ($10\pm5\%$ and $1.3\pm1.0\%$, respectively, where errors here are at 90\% confidence level).
An important result of our work here is that the temperature decline rate ($2.1\pm 0.2\%$ for constant $\NH$ and $2.7\pm 0.3\%$ for varying $\NH$) is significantly lower than values previously reported based on a smaller set of ACIS-S GRADED observations ($3.9\pm0.7\%$ per 10 years from \citealt{2010ApJ...719L.167H} and $3.5\pm0.4\%$ per 10 years from \citealt{2013ApJ...777...22E}).
Therefore, the rates of temperature decline measured using HRC-S data and ACIS-S subarray and GRADED data all agree at $\sim 90\%$ confidence level.

Very recently, \citet{2018ApJ...864..135P} follow up their previous analysis by including a new ACIS-S subarray observation, which expands this dataset to 3 measurements in 9 years, and use \textsc{CALDB} 4.7.3\footnote{\textsc{CALDB} 4.7.3 includes contaminant model N0010 for ACIS-S.  Similarly, \textsc{CALDB} 4.7.8 (used in Appendix~\ref{sec:appendix-casa}) includes N0010.  On the other hand, the most recent calibration \textsc{CALDB} 4.8.1 (used in Section~\ref{sec:casa}) includes the latest contaminant model N0012 for ACIS-S (see \url{http://cxc.cfa.harvard.edu/caldb4/downloads/Release\_notes/CALDB\_v4.8.1.html}). \label{foot:caldb}} (also testing versions up to 4.7.7); their analysis yields 3$\sigma$ upper limits on the ten-year rate of temperature decline of 2.4\% assuming a constant $\NH$ and 3.3\% assuming a varying $\NH$.
In our new analysis using 13 measurements in 18 years, as well as the latest \Chandra\ calibration and contamination models (\textsc{CALDB} 4.8.1; see footnote~\ref{foot:caldb} and Appendix~\ref{sec:appendix-casa}), the temperature decline we measured using ACIS-S GRADED mode is in agreement with the upper limits found by \cite{2018ApJ...864..135P} using the subarray observations.

As noted above, our spectral fits suggest $\NH$ may be variable (see Table~\ref{tab:casa} and Fig.~\ref{fig:nh}). \cite{2018ApJ...864..175A} study the effect on $\NH$ of material expelled by a supernova using three-dimensional supernova simulations. They find that different supernova models do not manifest measurable differences in $\NH$ at the current age of the Cassiopeia A neutron star. They also  analyse the same three Chandra observations (with $\NH$ tied across observations) as those used by \cite{2018ApJ...864..135P} and measure a ten-year rate of temperature decline of $\sim$1$\pm$1$\%$ and a supernova ejecta contribution to $\NH$ that is indistinguishable from that due to the intervening interstellar medium. While motion of intervening supernova remnant material (or the NS) as the cause of varying $\NH$ cannot be ruled out, a varying $\NH$ likely reflects incomplete modelling of the detector contaminant (see, e.g., \citealt{2018SPIE10699E..6BP}). We note however that in the case of our spectral fits allowing for a varying $\NH$, the ten-year rates of temperature and flux decline are actually larger, i.e., $2.7\pm0.3\%$ and $\approx 8\%$, respectively (see Fig.~\ref{fig:temp}), than the case when $\NH$ is constant.  Recent analysis by \citet{2018SPIE10699E..6BP} of the ACIS contaminant suggests the build up on the detector is not increasing as much as predicted by model N0010, and this effect is taken into account in the revised model N0012 (see footnote~\ref{foot:caldb}).  Future refinements of the ACIS-S calibration may aid attempts to more accurately measure the cooling rate of the NS in Cassiopeia~A.

\section*{Acknowledgements}

COH thanks B. Posselt for discussions. The authors are grateful to the anonymous referee for comments which improved the manuscript. WCGH acknowledges support through grant ST$/$R00045X$/$1 from Science and Technology Facilities Council in the UK.
COH is supported by NSERC Discovery Grant RGPIN-2016-04602 and a Discovery Accelerator Supplement.
MB is fully and DP partially supported by the Consejo Nacional de Ciencia y Tecnolog\'ia with a CB-2014-1 grant $\#$240512. This research made use of NumPy \citep{van2011numpy}, SciPy \citep{jones_scipy_2001} and matplotlib, a Python library for publication quality graphics \citep{Hunter:2007} .




\bibliographystyle{mnras}
\bibliography{references_v2} 




\appendix

\section{Analytic T$_s$-T$_b$ relations}
\label{sec:appendix_tbts}

In this section, we present accurate fits to the computed $\Ts$-$\Tb$-$\rho^*$ data which can be used in cooling simulations. All the fits are obtained for an envelope with surface gravity $g_{s,0}$ = 2.4271 $\times$ 10$^{14}$ cm s$^{-2}$ but can be scaled for any $g_s$ using $Y = (\Ts/1{\rm MK})(g_{s,0}/g_s)^{1/4}$ \citep{1983ApJ...272..286G}. Firstly, we obtain $\Tb$($\Ts$, $\rho^*$) fits and use an adapted version of the shape of the analytic functions (Equation \ref{eq:fiteq}) presented by \cite{2016MNRAS.459.1569B}. 

\begin{equation}
\begin{split}
\label{eq:fiteq}
T_b (Y, \rho*) = 10^7 K \times \Biggl( f_4(Y) + [f_1(Y) - f_4(Y)] \\
\times \left[f_5(Y, \rho*) + \left( \frac{\rho*}{f_2(Y)}\right)^{f_3(Y)}\right]^{p_{12}} \Biggr)
\end{split}
\end{equation}

\subsection{H-He mixture}

The analytic results for the H-He mixture are applicable to an envelope with a lower boundary of $\rho_b$ = 10$^{8}$ g cm$^{-3}$. While the H-He temperature relations can be fit with a simple function needing less fit parameters, we fit the same function as \cite{2016MNRAS.459.1569B} for consistency. The best fit parameters correspond to a maximum relative error of 0.0054 and a rms of the relative error of 0.0028.

\begin{equation}
\begin{split}
\label{eq:fitsubeq-HHe}
f_1(Y) &= p_1Y^{p_2} \sqrt{1 + p_3 Y^{p_4}} \\
f_2(Y) &= p_5 Y^{p_6} \sqrt{1 + p_7 Y^{p_8}} \\
f_2(Y) &= \frac{p_9 Y^{p_10}}{\left(1 - p_{11} Y + p_{12} Y^{2}\right)^2} \\
f_3(Y) &= p_{13} Y^{-p_{14}} \\
f_{5}(Y) &= 1
\end{split}
\end{equation}

\begin{center}
	\begin{table*}
		\caption{Best fit parameters for the H-He mixture with $\rho_b$ = 10$^{8}$ g cm$^{-3}$ (Equations~\ref{eq:fitsubeq-HHe}).}
		\label{tab:fitparams-HHe}
		\begin{tabular}{c c c c c c c c} 
			\hline
			$p_1$ & $p_2$ & $p_3$ & $p_4$ & $p_5$ & $p_6$ & $p_7$ & $p_8$ \\
			\hline
			0.8254 & 2.086 & 18.70 & -1.025 & 3.346 & 1.589 & 0.03848 & 1.572  \\
		\end{tabular}
		
		\begin{tabular}{c c c c c c c}
			\\
			$p_9$ & $p_{10}$ & $p_{11}$ & $p_{12}$ & $p_{13}$ & $p_{14}$ & $p_{15}$ \\
			\hline
			21.03 & 3.268 & 0.2416 & 0.1923 & -1.355 & -0.07620 & -9.864 $\times$ 10$^{3}$ \\
			\hline
		\end{tabular}	
		
	\end{table*}
\end{center}

\subsection{He-C mixture}

The analytic results for the He-C mixture for an envelope with a lower boundary of $\rho_b$ = 10$^{10}$ g cm$^{-3}$. Equation \ref{eq:fiteq} consists of the functions given in \ref{eq:fitsubeq}. Here, the functions $f_1-f_4$ are identical to those in \cite{2016MNRAS.459.1569B}. The function $f_5$ is an additional term to accurately fit the DNB models. For the He-C mixture, we obtain a maximum relative error of 0.019 which is slightly larger than the maximum relative error of 0.017 that \cite{2016MNRAS.459.1569B} obtain for the C-Fe envelope. The root mean square of the relative error is 0.0037.


\begin{equation}
\begin{split}
\label{eq:fitsubeq}
f_1(Y) &= p_1Y^{p_2 \text{log}_{10}Y+p_3} \\
f_2(Y) &= p_7 Y^{p_8 (\text{log}_{10} Y)^2 + p_9} \\
f_3(Y) &= p_{10} \sqrt{\frac{Y}{Y^2 + p_{11}^2}} \\
f_4(Y) &= p_4 Y^{p_5 \text{log}_{10} Y + p_6} \\
f_{5}(Y, \rho*)& = p_{13} Y^{p_{14}} + p_{15} \rho*^{p_{16}}
\end{split}
\end{equation}

\begin{center}
\begin{table*}
	\caption{Best fit parameters for the He-C mixture with $\rho_b$ = 10$^{10}$ g cm$^{-3}$  (Equations~\ref{eq:fitsubeq}).}
	\label{tab:fitparams-HeC}
	\begin{tabular}{c c c c c c c c} 
		\hline
		$p_1$ & $p_2$ & $p_3$ & $p_4$ & $p_5$ & $p_6$ & $p_7$ & $p_8$ \\
		\hline
		10.927 & 0.977 & 2.925 & 3.521 & 0.0346 & 1.650 & 963.474 & -19.318  \\
		\\
		$p_9$ & $p_{10}$ & $p_{11}$ & $p_{12}$ & $p_{13}$ & $p_{14}$ & $p_{15}$ & $p_{16}$ \\
		\hline
		-0.846 & 2.931 &1.190 & -0.175 & 3.110 $\times$ 10$^{3}$ & 10.812 & 2.181 $\times$ 10$^{-9}$ & 2.314 \\
		\hline
	\end{tabular}	
\end{table*}
\end{center}

\section{Analysis of Cassiopeia~A data using previous \Chandra\ contamination model N0010}
\label{sec:appendix-casa}

Here we show results of our analysis of the same data as that in Section~\ref{sec:casa} but using a previous \textsc{CALDB} (4.7.8 in \textsc{CIAO} 4.9), with ACIS-S contamination model N0010, and \textsc{Xspec} (12.9.1).  The results presented in this Appendix are to illustrate the small changes that the ACIS-S contamination model N0012 (in \textsc{CALDB 4.8.1}) introduces, as well as to facilitate comparisons with the results of \citet{2018ApJ...864..135P} who use \textsc{CALDB} 4.7.3 and N0010.  The analysis follows the same procedure as that described in Section~\ref{sec:casa}, and the final fit results are given in Table~\ref{tab:casa_old}.  Comparison with the fit results in Table~\ref{tab:casa} show differences in temperature and flux well within uncertainties, keeping in mind that some of the small changes are due to differing best-fit radius (13.0~km versus 12.9~km) and hence gravitational redshift.

\begin{table*}
	\caption{
		Surface temperature $\Tssix$ ($10^{6}\mbox{ K}$), absorbed 0.5--7~keV flux $F_{\rm -13}^{\rm abs}$
		($10^{-13}\mbox{ erg cm$^{-2}$ s$^{-1}$}$), and absorption column
		$\NHtwo$ ($10^{22}\mbox{ cm$^{-2}$}$) determined from model fits using \textsc{CALDB} 4.7.8 (in comparison to Table~\ref{tab:casa} which results from using \textsc{CALDB} 4.8.1) to
		\Chandra\ ACIS-S GRADED spectra of the neutron star in the Cassiopeia~A
		supernova remnant.
		Three spectral fit results are shown, two with constant $\NH$ and one with
		changing $\NH$.
		Each set of 13 temperatures and fluxes are fit to a linear decline, with
		decline rate and fit statistic as shown.
		For merged ObsIDs, the MJD listed is that of the first ObsID.
		Number in parentheses is $1\sigma$ uncertainty in last digit.
	}
	\label{tab:casa_old}
	\begin{tabular}{rlcccccccc}
		\hline
		& & $\NHtwo$ & \multicolumn{2}{c}{1.73} & \multicolumn{2}{c}{1.68} & \multicolumn{3}{c}{see below} \\
		& & $M$ ($\Msun$) & \multicolumn{2}{c}{1.65} & \multicolumn{2}{c}{1.65} & \multicolumn{3}{c}{1.65} \\
		& & $R$ (km) & \multicolumn{2}{c}{10.3} & \multicolumn{2}{c}{13.0} & \multicolumn{3}{c}{13.0} \\
		& & $\chi^2$/dof & \multicolumn{2}{c}{1550/1450} & \multicolumn{2}{c}{1512/1450} & \multicolumn{3}{c}{1486/1437} \\ \\
		ObsID & Date & MJD & $\Tssix$ & $F_{\rm -13}^{\rm abs}$ & $\Tssix$ & $F_{\rm -13}^{\rm abs}$ & $\Tssix$ & $F_{\rm -13}^{\rm abs}$ & $\NHtwo$ \\
		\hline
		114 & 2000 Jan 30 & 51573.4 & 2.127(10) & 7.4(2) & 1.862(8) & 7.3(2) & 1.870(10) & 7.3(2) & 1.73(3) \\
		1952 & 2002 Feb 6 & 52311.3 & 2.125(10) & 7.4(2) & 1.861(8) & 7.3(2) & 1.869(10) & 7.3(2) & 1.73(3) \\
		5196 & 2004 Feb 8 & 53043.7 & 2.106(10) & 7.1(1) & 1.845(8) & 7.0(2) & 1.842(10) & 7.0(1) & 1.66(3) \\
		9117/9773 & 2007 Dec 5/8 & 54439.9 & 2.097(9) & 7.0(2) & 1.835(8) & 6.9(2) & 1.842(10) & 6.9(2) & 1.72(4) \\
		10935/12020 & 2009 Nov 2/3 & 55137.9 & 2.089(10) & 6.9(2) & 1.828(8) & 6.8(2) & 1.832(11) & 6.8(2) & 1.71(4) \\
		10936/13177 & 2010 Oct 31/Nov 2 & 55500.2 & 2.080(10) & 6.7(1) & 1.820(8) & 6.6(2) & 1.812(11) & 6.6(1) & 1.64(4) \\
		14229 & 2012 May 15 & 56062.4 & 2.044(9) & 6.4(2) & 1.791(7) & 6.3(1) & 1.797(11) & 6.3(2) & 1.72(4) \\
		14480 & 2013 May 20 & 56432.6 & 2.062(9) & 6.5(1) & 1.806(8) & 6.4(2) & 1.810(10) & 6.5(1) & 1.70(4) \\
		14481 & 2014 May 12 & 56789.1 & 2.045(9) & 6.3(1) & 1.792(7) & 6.2(2) & 1.799(10) & 6.2(2) & 1.72(4) \\
		14482 & 2015 Apr 30 & 57142.5 & 2.064(9) & 6.6(2) & 1.809(7) & 6.6(2) & 1.795(10) & 6.5(2) & 1.60(4) \\
		19903/18344 & 2016 Oct 20/21 & 57681.2 & 2.059(9) & 6.6(2) & 1.803(8) & 6.5(1) & 1.789(11) & 6.4(1) & 1.60(4) \\
		19604 & 2017 May 16 & 57889.7 & 2.050(9) & 6.5(2) & 1.798(7) & 6.4(2) & 1.795(10) & 6.4(1) & 1.67(4) \\
		19605 & 2018 May 15 & 58253.7 & 2.050(9) & 6.5(2) & 1.798(7) & 6.5(2) & 1.777(10) & 6.4(1) & 1.53(4) \\
		\hline
		& & 10-year & & & & & & & \\
		& & decline rate & $2.2\pm0.2\%$ & $7.3\pm1.0\%$ & $2.1\pm0.2\%$ & $7.2\pm1.0\%$ & $2.6\pm0.3\%$ & $7.6\pm0.9\%$ & \\
		& & $\chi^2$/dof & 15.8/11 & 12.9/11 & 16.8/11 & 15.4/11 & 6.3/11 & 11.1/11 & \\
		\hline
	\end{tabular}
\end{table*}


\bsp	
\label{lastpage}
\end{document}